\documentclass[12pt]{article}
\pdfoutput=1
\usepackage[square,comma,numbers,sort&compress]{natbib}
\usepackage{graphicx,epstopdf,amssymb,amsfonts,amsmath,amsthm,array,
mathrsfs,amscd}
\usepackage[vcentermath]{youngtab}
\usepackage{float}
\usepackage{hyperref}
\newcommand{\pbs}[1]{\let\temp=\\#1\let\\=\temp}
\numberwithin{equation}{section}
\def\be{\begin{equation}}\def\ee{\end{equation}}
\def\cvp{\raise 2pt\hbox{,}} 
 \def\tr{\mathop{\rm tr}\nolimits}

 \def\d{{\rm d}}\def\nn{{\cal
N}} 
\def\suN{\text{SU}(N)} \def\uN{\text{U}(N)}\def\uK{\text{U}(K)}

\def\gs{g_{\text s}}\def\ls{\ell_{\text s}}
\def\la{\lambda}\def\La{\Lambda}
\def\DBI{S_{\text{DBI}}}\def\CS{S_{\text{CS}}}
\def\EH{S_{\text{EH}}}
\def\Sg{S_{\text g}}\def\Sp{S_{\text p}}\def\Sct{S_{\text{c.t.}}}
\def\AdS{\text{AdS}}

\def\CT{S_{\text{c.t.}}}
\def\rh{r_{\text{h}}}\def\uh{u_{\text{h}}}
\def\uh{u_{\text{h}}}

\def\M{M}
\def\P{\text{P}}
\def\Si{\Sigma}
\def\del{\partial}
\def\GN{G_{\text{N}}}
\def\Vol{\text{Vol}}

\def\Mtwo{\tau_{{\text{M2}}}}
\def\Mfive{\tau_{{\text{M5}}}}

\def\alp{\alpha'}
\def\Sugra{S_{\text{sugra}}}
\def\ve{\eta}
\def\Re{\text{Re}}
\def\R{{\cal R}}
\def\Rd{{\mathscr  R}}
\def\LPM{\ell_{11}}

\def\Sfive{S_{5}}
\def\plb#1#2#3{{\it Phys.\ Lett.\ }{\bf B #1} (#2) #3}
\def\npb#1#2#3{{\it Nucl.\ Phys.\ }{\bf B #1} (#2) #3}

\def\prl#1#2#3{{\it Phys.\ Rev.\ Lett.\ }{\bf #1} (#2) #3}
\def\jhep#1#2#3{{\it J. High Energy Phys.\ }{\bf #1} (#2) #3}
\def\prd#1#2#3{{\it Phys.\ Rev.\ }{\bf D #1} (#2) #3}

\def\atmp#1#2#3{{\it Adv.\ Theor.\ Math.\ Phys.\ }{\bf #1} (#2) #3}
\def\cmp#1#2#3{{\it Comm.\ Math.\ Phys.\ }{\bf #1} (#2) #3}

\def\ijmpa#1#2#3{{\it Int.\ J.\ Mod.\ Phys.\ }{\bf A #1} (#2) #3}
\def\mpla#1#2#3{{\it Mod.\ Phys.\ Lett.\ }{\bf A #1} (#2) #3}

\def\cqg#1#2#3{{\it Class.\ Quant.\ Grav.\ }{\bf #1} (#2) #3}

\def\rpp#1#2#3{{\it Rept.\ Prog.\ Phys.\ }{\bf #1} (#2) #3}
\def\imath#1#2#3{{\it Invent math }{\bf #1} (#2) #3}

\def\cag#1#2#3{{\it Comm.\ Anal.\ Geom.\ }{\bf #1} (#2) #3}

\begin{document}
%
%
{\pagestyle{empty}
\begin{flushright} MIT-CTP/4770\\
CERN-TH-2016-038
\end{flushright}
\parskip 0in
\

\vfill

\begin{center}
%
%

{\LARGE Gravity and On-Shell Probe Actions}

\vspace{0.4in}

Frank F{\scshape errari}$^{* \dagger}$ and Antonin R{\scshape ovai}$^{\star *\diamond\ddagger}$
\\

\medskip

$^{*}$\textit{Universit\'e libre de Bruxelles (ULB) and International Solvay Institutes\\
Service de Physique Th\'eorique et Math\'ematique\\
Campus de la Plaine, CP 231, B-1050 Bruxelles, Belgique}

\smallskip

$^{\star}$\textit{D\'epartement de Physique Th\'eorique\\
Universit\'e de Gen\`eve\\
24, quai Ansermet, CH-1211 Gen\`eve 4, Suisse}

\smallskip

$^{\dagger}$\textit{Theoretical Physics Department\\
CERN, CH-1211 Gen\`eve, Suisse}

\smallskip

$^{\diamond}$\textit{Center for Theoretical Physics\\
Massachusetts Institute of Technology\\
77 Massachusetts Avenue\\
Cambridge, MA 02139, USA}

\smallskip

$^{\ddagger}$\textit{Arnold Sommerfeld Center for Theoretical Physics\\
Ludwig-Maximilians-Universit\"at M\"unchen\\
Theresienstrasse 37, D-80333 M\"unchen, Deutschland
}

\smallskip
\texttt{frank.ferrari@cern.ch, antonin.rovai@unige.ch}
\end{center}
\vfill\noindent

In any gravitational theory and in a wide class of background space-times, we argue that there exists a simple, yet profound, relation between the on-shell Euclidean gravitational action and the on-shell Euclidean action of probes. The probes can be, for instance, charged particles or branes. The relation is tightly related to the thermodynamic nature of gravity. We provide precise checks of the relation in several examples, which include both asymptotically flat and asymptotically AdS space-times, with particle, D-brane and M-brane probes. Perfect consistency is found in all cases, including in a highly non-trivial example including $\alpha'$-corrections.


\vfill

\medskip
%
\begin{flushleft}
\today
\end{flushleft}
%
\newpage\pagestyle{plain}
\baselineskip 16pt
\setcounter{footnote}{0}

}

\tableofcontents

\newpage
\section{\label{IntroSec} Introduction}

Gravitation is both the most universal and less understood of all the fundamental forces. Many decades of theoretical research lead to the idea that its profound nature may be entirely different from the one of the other forces, which can be described by canonically quantizing some classical field theories. Instead, the usual ``classical'' Einstein's theory of gravity, together with the notion of a ``classical'' space-time, seem to be emerging in a thermodynamic limit where the number of degrees of freedom of the underlying fundamental quantum description becomes very large.

Serious arguments in favour of this point of view date back to papers by Bekenstein and Hawking in the 70s (see e.g.\ \cite{Bek1,BCH,Haw1,Jac,Padma} and references therein). Much stronger evidence comes from explicit constructions in string theory in the context of Maldacena's holographic correspondence \cite{malda,shol}. In this set-up, the classical, thermodynamic-like limit of gravity is typically related to the 't~Hooft large $N$ limit of gauge theories. However, even in this context where the gravitational theory is in principle precisely defined in terms of an ordinary quantum system, it is hard to understand how the classical bulk space-time and gravity emerge. In particular, space-time locality and the Equivalence Principle remain to a large extent mysterious. Any progress in this direction will undoubtedly have profound and surprising consequences in our understanding of gravitation. Interesting trails have been vigorously explored in recent years, in particular in black hole physics and in relation with the holographic description of entanglement (see e.g.\ \cite{GenRef} and references therein).

The aim of the present paper is to explain and study a simple but rather surprising and profound relation in gravity which, as we shall argue, is directly related to the thermodynamic nature of the theory. The relation makes a link between two important but seemingly unrelated objects. On the one hand, we have the on-shell Euclidean gravitational action, denoted by $\Sg^{*}$. It has been known for a long time to play a central role in the study of the thermodynamics of black holes \cite{GHawk} and, in holography, it is associated with the generating functional of the boundary planar correlation functions \cite{shol}. On the other hand, we have the on-shell Euclidean action of probes of the background geometry, denoted by $\Sp^{*}$. The probes can be, for example, charged particles or branes.

A special instance of the relation between $\Sg^{*}$ and $\Sp^{*}$, whose general form will be presented in the next section, was first found in \cite{fer1}, by studying in details the construction of probe D-brane actions from the field theory point of view.\footnote{The main goal of \cite{fer1} was to explain how the holographic space-time can be explicitly seen to emerge from gauge theory, using the notion of D-brane probes. See also \cite{ferrelated} for related works. Some relevant aspects of Ref.\ \cite{fer1} are briefly reviewed in Appendix B.} The relation was then further explored in \cite{FerRov} from the point of view of the bulk gravitational theory, in the context of the holographic correspondence, for a general class of asymptotically AdS space-times. In all these cases, full consistency was found, by using in particular a non-trivial isoperimetric inequality (this will be reviewed in Section \ref{Sec3}). This isoperimetric inequality was also studied thoroughly in an interesting recent paper \cite{McInnes}, in a very wide range of backgrounds going beyond the space-times considered in \cite{FerRov}. Consistency is found in all cases, which leads the authors of \cite{McInnes} to propose that our relation\footnote{More precisely, the authors of \cite{McInnes} focus on the isoperimetric inequality, which is directly related to the fundamental relation in some special instances, as explained in \cite{FerRov} (see also Section \ref{Sec3}).} should be considered to be a crucial requirement in any gravitational theory; in the wording of \cite{McInnes}, a ``law of physics'' coming from the mathematical consistency of the underlying theory. Related and extremely instructive results using probe branes in holographic set-ups have also appeared in the earlier literature \cite{KTR}.

One of the main point of the present paper will be to argue that, indeed, the relation between $\Sg^{*}$ and $\Sp^{*}$ should be considered to be a basic property of any theory of gravitation, its universality being tightly related to the thermodynamic nature of gravity.\footnote{Except for extensions discussed in the conclusion section, we shall always consider gravity in the strict thermodynamic (i.e.\ semi-classical) limit.} This universality goes far beyond the examples considered in \cite{fer1, FerRov, McInnes}. In particular, it includes cases with higher derivative corrections to the usual Einstein theory, and it also works in asymptotically flat space-times. Explicit evidence will be provided below.

The plan of the paper is as follows. 

In Section \ref{Sec2}, we derive the precise form of the general relation between $\Sg^{*}$ and $\Sp^{*}$ using a simple argument. We then discuss the case of a probe charged particle in the asymptotically flat Reissner-Nordstr\"om geometry. This allows to illustrate, in a very simple example, the main basic features, including the important issue of the precise definition of the probe action. We also treat the case of a D3-brane probe in the asymptotically flat black D3-brane geometry, which is very similar.

In Section \ref{Sec3}, we focus on the case of probe branes in holographic set-ups. We first show that, in many interesting cases, our general relation becomes a simple proportionality between $\Sg^{*}$ and $\Sp^{*}$, with a fixed proportionality factor depending on the particular brane system under study. We then consider the standard near-horizon limit of the black D3-brane system studied in Section \ref{Sec2} and show that $\Sg^{*}$ and $\Sp^{*}$ are indeed proportional in this limit. Using the string theoretic formulas relating the D3-brane tension and the five dimensional Newton constant, we find that the proportionality factor precisely matches the one predicted by our fundamental relation. The discussion of the near-horizon limit also allows to illuminate some subtle and crucial issues, first discussed in \cite{FerRov}, associated with the precise definition of the probe brane action in asymptotically AdS spaces. We then turn to the more general case of a $(d-1)$-brane probing an asymptotically $\text{AdS}_{d+1}$ space with an arbitrary boundary metric.\footnote{More precisely, the Yamabe constant of the boundary metric must be non-negative. This technical condition is required for the consistency of the boundary theory \cite{WY}.} For completeness, we briefly reproduce the analysis in \cite{FerRov} showing that our relation between $\Sg^{*}$ and $\Sp^{*}$ follows from an interesting isoperimetric inequality first derived in \cite{Wang}. We also discuss the case of other D-brane and M-brane systems, checking in particular that the proportionality factors between $\Sg^{*}$ and $\Sp^{*}$ predicted by our relation always match the factor computed by using the string theoretic formulas for the various brane tensions and Newton's constants. 

In Section \ref{alphaSec}, we turn to an example for which higher derivative, $\alpha'$-corrections to the gravitational theory are taken into account. First, we show that the general relation between $\Sg^{*}$ and $\Sp^{*}$ predicts a correction to the simple proportionality law used in Section \ref{Sec3} when $\alpha'$-corrections are included. We then check explicitly the validity of the resulting formula for the case of the $\alpha'$-corrected Schwarzschild-$\text{AdS}_{5}$ background. This provides an extremely non-trivial test of our ideas.

We have also included a conclusion section and several technical appendices complementing the main text.

\section{\label{Sec2} The fundamental relation and two examples}
\subsection{\label{DerSubSec} Derivation}

The general philosophy, from which we start, is that any solution\footnote{In practice, we limit ourselves to static backgrounds in the present paper.} $\mathcal B$ in a well-defined semi-classical gravitational theory corresponds to a state $|\mathcal B\rangle$ in some thermodynamic limit of an ordinary\footnote{Ordinary here means a standard non-mysterious quantum mechanical system with a well-defined Hermitian Hamiltonian $H$. In particular, this system is not a quantum version of classical gravity.} quantum mechanical system. This situation is of course realized in the usual holographic correspondence, but, more generally, we assume that it should be valid in any consistent formulation of gravitation. The correspondence means that there exists a dictionary between the observables (Hermitian operators) of the ordinary quantum system, averaged in the state $|\mathcal B\rangle$, and the diffeomorphism invariant observables of the gravitational theory, evaluated on the background $\mathcal B$. Of course, this dictionary can be subtle and is often only partially known, even in standard holographic set-ups. 

For our purposes, the observable we focus on in the gravitational theory is the on-shell Euclidean gravitational action $\Sg^{*}$, obtained by evaluating the gravitational action $\Sg$ on the background $\mathcal B$. The action $\Sg$ could be the usual Einstein-Hilbert plus Gibbons-Hawking action, or any consistent generalization, involving for example supersymmetry and/or higher derivative terms. The relevance of $\Sg^{*}$ has been known for a long time \cite{GHawk}. Its precise definition requires some care, since its naive value is usually infinite due to the non-compactness of space-time. In the holographic set-up, we use the holographic renormalization procedure to make sense of it \cite{holren}.

The correspondence between the gravitational theory and the ordinary quantum system implies that $\Sg^{*}$ coincides with some physical quantity in the quantum theory, when the appropriate thermodynamic limit is taken. For example, when one considers a black hole background, one has the famous relation
\be\label{BHFree} e^{-\Sg^{*}}=Z\ee
relating $\Sg^{*}$ to the partition function $Z=\tr e^{-\beta H}$ of the underlying quantum mechanical system, computed at a temperature $T=1/\beta$ coinciding with the Hawking temperature of the black hole. The relation \eqref{BHFree} allows to derive in a very neat way the thermodynamic properties of black holes, in both asymptotically flat and asymptotically AdS space-times (see e.g.\ \cite{GHawk,HP}). The free energy is $F=T\Sg^{*}$ and the energy and entropy are obtained by using the standard thermodynamic identities. Note that $Z$, or equivalently $F$, always depends on $T$, but may also depend on a set of conserved charges. Since we always work in the thermodynamic limit, the distinction between the canonical and grand canonical description is immaterial.

More generally, $e^{-\Sg^{*}}$ yields the generating functional $Z$ of planar correlation functions in the holographic dictionary \cite{shol}. In this context, the quantum mechanical system is typically a boundary CFT, obtained by taking the low energy limit of the worldvolume dynamics of a stack of branes. For example, the maximally supersymmetric Yang-Mills theory with gauge group $\uN$ is obtained from a stack of $N$ D3-branes. The generating functional $Z$ thus depends on $N$, or more generally on a set of integers. These integers can be treated in strict parallel with the ordinary conserved charges. Actually, in the string theory constructions, they count the number of branes, which are ordinary charged objects.

We denote by $Q$ one of the conserved charges. It is always taken to be very large in microscopic units, in order to be in the thermodynamic limit. In the quantum mechanical system, we can compute $Z(Q)$. In the gravitational description, we have a Euclidean background space-time $\mathcal B_{Q}$. The gravitational action evaluated on this space-time is $\Sg^{*}(Q)=-\ln Z(Q)$. Let us now imagine that we deform the system in such a way that $Q$ is changed to $Q+q$, with $|q|\ll |Q|$. We are going to assume that, at least in a wide variety of cases, this small deformation has two natural descriptions in the gravitational theory:

i) An obvious description is simply to deform the background space-time from $\mathcal B_{Q}$ to $\mathcal B_{Q+q}$. This of course yields
\be\label{Zdef1} -\ln Z(Q+q) = \Sg^{*}(Q+q) = \Sg^{*}(Q) + q\frac{\partial\Sg^{*}}{\partial Q}(Q)\, ,\ee
where, in the second equality, we have used the fact that the deformation is small.

ii) A second natural description is to keep the background $\mathcal B_{Q}$ undeformed, but to add a probe object (particle or brane) of charge $q$ in $\mathcal B_{Q}$. The probe object has an action $\Sp$, typically the sum of a kinetic term and a coupling to the gauge potential associated with the charge. The total action is $\Sg + \Sp$. We assume that the semi-classical limit is still valid and thus, in this second description, we get
\be\label{newZ1} -\ln Z(Q+q)=\Sg^{*}(Q)+\Sp^{*}\, ,\ee
where $\Sp^{*}$ is the on-shell (minimal) value of the action $\Sp$ in the undeformed background $\mathcal B_{Q}$. To minimize $\Sp$, all possible probe worldvolumes in the bulk must be considered, without any particular boundary condition.\footnote{Note that this is very different from other contexts where one considers the minimization of brane actions with specific boundary conditions at infinity, like for instance in the holographic computation of Wilson loops. However, our ideas can also be applied in such set-ups, see Section 5 for a brief discussion.} The only geometrical constraint is that the worldvolume must span the time direction. Note that the fact that the charged object is a probe implies that $|\Sp^{*}|\ll |\Sg^{*}|$, and the fact that the semi-classical limit is valid implies that $|\Sp^{*}|$ is still very large in microscopic units.\footnote{In the cases where the gravitational theory is understood beyond the semi-classical limit, these assumptions can be waived, at least in principle.} 

Comparing \eqref{Zdef1} and \eqref{newZ1}, we get our fundamental relation between the on-shell gravitational action and the on-shell action of a probe,
\be\label{fundrel} \boxed{\Sp^{*} = q\frac{\partial\Sg^{*}}{\partial Q}\,\cdotp}\ee
Several comments are here in order.

a) Clearly, the second point of view above, which yields \eqref{newZ1}, does not constitute a rigorous proof. Such a proof may only be given in set-ups where the quantum mechanical system and the dictionary with the gravitational description are fully known. However, we do expect its range of validity to be large, with only natural conditions to be imposed on the charged probes (a typical condition is, for example, a BPS bound $|m|\geq |q|$, see below). The universality of \eqref{fundrel} is rooted in the thermodynamic nature of gravity, the fact that only a few macroscopic features, like the total charge $Q$, determine the background.

b) Eq.\ \eqref{fundrel} implies that terms of order $O(q^{0})$, typically the kinetic terms, vanish in $\Sp^{*}$. This is a simple but non-trivial requirement, that will turn out to be valid in all the examples studied below. Note that terms of order $O(q^{2})$ in $\Sp^{*}$, if not altogether absent, must be neglected since we are in the probe approximation. We can thus always write
\be\label{Spfact} \Sp^{*}=q \mathcal A^{*}(Q)\, ,\ee
where $\mathcal A^{*}=\partial\Sg^{*}/\partial Q$ does not depend on $q$.

c) The objects on the two sides of the equality \eqref{fundrel} look very different. In particular the gravitational action is a bulk quantity whereas the probe action is computed along a worldline or worldvolume. This makes the relation surprising and particularly interesting. For example, for a black hole of charge $Q$, \eqref{fundrel} and \eqref{Spfact} yield
\be\label{murel} \mu = T \mathcal A^{*}\ee
for the chemical potential of the black hole. This formula provides an entirely new and rather simple way to obtain the chemical potential. 
More generally, in many cases (see e.g.\ Section \ref{Sec3}), the full free energy (or gravitational action) can be obtained straightforwardly from $\Sp^{*}$ by integrating with respect to $Q$.

d) The cases where the probes are branes and the charge simply counts the number of branes is particularly interesting and will be discussed at length in the following sections. In these cases, it is often possible to justify \eqref{fundrel} using a slight modification of Maldacena's original argument for the AdS/CFT duality \cite{malda}. For example, consider $N$ D3-branes, $N\gg 1$. The Maldacena argument implies that a large number $N\gg 1$ of branes can be replaced by the $\text{AdS}_{5}\times\text{S}^{5}$ background without brane, the ratio between the AdS scale and the five-dimensional Planck length being $(L/\ell_{\text P})^{3}\sim N^{2}$. If we consider $N+1$ branes instead of $N$, then the same argument can be repeated either by considering all the $N+1$ branes together, which yields again the $\text{AdS}_{5}\times\text{S}^{5}$ background without brane but now with a slightly modified ratio $(L/\ell_{\text P})^{3}\sim (N+1)^{2}$, or by replacing only $N$ branes by the $\text{AdS}_{5}\times\text{S}^{5}$ background with undeformed ratio $(L/\ell_{\text P})^{3}\sim N^{2}$, the additional brane, moving in the background generated by all the other branes, being kept explicitly. The equivalence of these two points of view immediately yields \eqref{fundrel} (this argument first appeared in \cite{fer1}).

\subsection{\label{RNsubSec} The Reissner-Nordstr\"om black hole}

\subsubsection*{The set-up}

Let us now illustrate the relation \eqref{fundrel} in the case of the standard charged black hole in four dimensions. The metric and electromagnetic field strength in the Euclidean are
\begin{gather}\label{metricRN} \d s^{2} = \frac{\Delta}{r^{2}}\d t^{2} + \frac{r^{2}}{\Delta}\d r^{2} + r^{2}\d\Omega_{2}^{2}\, ,\\\label{A}
F = i\frac{Q}{r^{2}}\,\d t\wedge\d r\, ,\end{gather}
where
\begin{gather}\label{DeltaRN} \Delta = r^{2}-2Mr + Q^{2} = (r-r_{+})(r-r_{-})\, ,\\ 
\label{rpmRN} r_{\pm} = M\pm\sqrt{M^{2}-Q^{2}}\, .
\end{gather}
The parameters $M$ and $Q$ correspond to the mass and the electric charge of the black hole, respectively. They must satisfy the usual BPS condition $|M|\geq |Q|$. The metric $\d\Omega_{2}^{2}$ is the standard round metric on the two-sphere of unit radius. The Euclidean geometry has the usual shape of a cigar, smoothness at the tip of the cigar (which corresponds to the location of the black hole horizon in the Minkowskian version of the geometry) being ensured by the periodicity condition $t\equiv t+\beta$ on the Euclidean time coordinate, where
\be\label{HawkTRN} T = \frac{1}{\beta} = \frac{r_{+}-r_{-}}{4\pi r_{+}^{2}}\ee
is the Hawking temperature. Note that the full Euclidean geometry is spanned when $r_{+}\leq r<\infty$. In particular, there is no ``interior'' of the black hole in Euclidean signature.

Let us now consider a particle of mass $m$ and charge $q$ probing the black hole background. Its Euclidean action is given by
\be\label{probeARN} \Sp = m\oint\!\d s - iq\oint\! A\, ,\ee
where the integral is taken along an arbitrary worldline wrapping the time circle, $\d s$ is the infinitesimal length (or Euclidean proper time) along the worldline and $A$ a gauge potential such that
\be\label{FdA} F=\d A\, .\ee
Our goal is to compute the chemical potential of the black hole from \eqref{fundrel}, or equivalently \eqref{murel}. We thus have to find the minimal value of the action \eqref{probeARN}, over all worldlines parameterized by $\beta$-periodic functions $r(t)$, $\theta(t)$ and $\phi(t)$, if $\theta$ and $\phi$ are the usual spherical angles over the $\text{S}^{2}$ part of the geometry. 

\subsubsection*{The precise definition of the probe action}

Before we discuss the minimization problem itself, which will be elementary, we have to address a crucial question regarding the precise definition of the probe action $\Sp$. The subtlety comes from the fact that the action \eqref{probeARN} is expressed in terms of the gauge potential $A$, not in terms of the gauge invariant field strength. This is irrelevant for the equations of motion, or for the variations of the action, which are expressed in terms of $F$ only. However, our fundamental relation \eqref{fundrel} involves the actual value of the action. As we shall easily discover, to fix this value unambiguously, one must have a precise prescription to pick a particular gauge potential $A$.

To understand the problem, let us consider the family of gauge potentials
\be\label{gpRN} A_{c} = i\Bigl(\frac{Q}{r}+c\Bigr)\d t\, ,\ee
labeled by an arbitrary constant $c$.\footnote{This is nothing but the usual arbitrary constant one may add to the electrostatic potential.} All these gauge potentials yield the correct field strength \eqref{A}, $F=\d A_{c}$. However, their contributions to the action \eqref{probeARN} differ by an additive constant depending on $c$,
\be\label{actcont} -iq\oint\! A_{c} = qQ\int_{0}^{\beta}\frac{\d t}{r(t)} + qc\beta\, .\ee
This undetermined constant\footnote{Constant here means that it does not depend on the worldline, but of course it may depend on the other parameters in the problem, like the charge or the temperature.} $qc\beta$ of course crucially affects the minimal value of the action. More generally, the field strength remains unchanged if one performs a gauge transformation 
\be\label{gtgen} A\mapsto A + \omega\, ,\ee
where $\omega$ is a priori an arbitrary closed one-form. The resulting ambiguity in the action is a term $iq\oint\omega$ which, by Stokes' theorem, does not depend on the worldline. The ambiguity associated with the general transformations \eqref{gtgen} is thus again a worldline-independent constant, which can change crucially the value of the minimum of the action.

One can think of two natural proposals to fix the above ambiguity. We are going to discuss them both, including the incorrect prescription, since this is a very important point that must be fully clarified.\footnote{We also want to discuss the incorrect proposal because it has been suggested to us on several occasions.}

\smallskip

\noindent\emph{Incorrect proposal}: impose that the gauge potential $A$ entering the probe action \eqref{probeARN} must be globally well-defined. 

This will clearly fix the ambiguity, at least in the Reissner-Nordstr\"om space-time we consider presently, since a globally defined gauge transformation $\omega$ must be exact and thus cannot change the action, $\oint\omega = 0$. If we consider the family of gauge potentials \eqref{gpRN}, they are not globally defined for generic values of $c$, due to the singular nature of the angular coordinate $t$ at the tip of the cigar $r=r_{+}$ (this is the same singularity that one encounters for the one-form $\d\theta$ at the origin of ordinary polar coordinates ($\rho$, $\theta$)). Regularity of $A_{c}$ at $r=r_{+}$ implies that $c=-Q/r_{+}$, yielding the globally smooth gauge potential
\be\label{Asmooth} A_{\text{smooth}}= iQ\Bigl(\frac{1}{r}-\frac{1}{r_{+}}\Bigr)\d t\ee
and the associated action $S_{\text{p, smooth}}$.

This proposal of global smoothness of the gauge potential might seem reasonable. For example, the smooth gauge potential \eqref{Asmooth} is the one that must enter into the definition of the Polyakov loop observable, since otherwise the loop would not be regular. However, it is obvious that the probe action is perfectly well-defined and regular for all values of the constant $c$ in \eqref{gpRN}. No regularity condition can fix $c$ in this case. The same is true for the field strength,\footnote{One can easily check that it is proportional to the area form of the cigar at the horizon $r=r_{+}$, which is smooth thanks to the choice \eqref{HawkTRN}.} that enters the equations of motion derived from the action and which is always globally well-defined.

\smallskip

\noindent\emph{Correct proposal}: impose that the probe action goes to the usual action $m\oint\d s$ for a point particle of mass $m$ in the asymptotically flat region of the geometry.

The action for a probe particle at a fixed position $(r,\theta,\phi)$, computed with the gauge potential \eqref{gpRN}, takes the form $\Sp = \beta V(r)$, where the ``potential energy'' $V_{c}$ is given by
\be\label{VRNex} V_{c}(r) = m\sqrt{1-\frac{2M}{r} + \frac{Q}{r^{2}}} + \frac{qQ}{r} + cq\, .\ee
The condition we propose to fix $c$ thus simply amounts to imposing that the potential energy reduces to the rest mass of the particle when it is infinitely far from the black hole. This is an extremely natural physical condition, we believe the only consistent and meaningful condition one can impose in an asymptotically flat background. It implies that
\be\label{ciszero} c=0\, .\ee
The correct gauge potential that must be used to compute the action is thus singular at $r=r_{+}$.

\subsubsection*{The minimum of the action}

Finding the minimum of the probe action is now very simple. By denoting by $\dot r = \d r/\d t$, etc., one first notes that
\begin{align}\label{ineqac1} \Sp &= \int_{0}^{\beta}\biggl[m\sqrt{\frac{\Delta}{r^{2}} + \frac{r^{2}}{\Delta}\dot r^{2}+r^{2}\bigl(\dot\theta^{2}+\sin^{2}\theta\,\dot\phi^{2}\bigr)} + \frac{qQ}{r}\biggr]\d t\\\label{ineqac2} 
& \geq\tilde\Sp = \int_{0}^{\beta}\! V_{0}(r)\,\d t\, .
\end{align}
The minimum of the action $\tilde\Sp$ is obtained for a worldline at a fixed position $r$ minimizing the potential $V_{0}$. Since, for such a worldline, $\tilde\Sp=\Sp$, this also yields the minimum of $\Sp$. Assuming that the probe satisfies the BPS bound $m\geq |q|$ (consistently with the BPS bound $M\geq |Q|$ satisfied by the black hole itself), it is trivial to check from \eqref{VRNex} that the minimum of $V_{0}$ is obtained for $r=r_{+}$. The on-shell probe action is thus
\be\label{onshellRN} \Sp^{*} = \beta V_{0}(r_{+}) = \beta\frac{qQ}{r_{+}}\,\cdotp\ee
Our fundamental relations \eqref{fundrel}, \eqref{Spfact}, \eqref{murel} thus yield the black hole chemical potential
\be\label{muRN}\mu = \frac{Q}{r_{+}}\,\cdot\ee
This is indeed the well-known correct value!\footnote{For completeness, it is interesting to recall how $\mu$ is traditionally computed. One evaluates the on-shell gravitational action $\Sg^{*}$ (Einstein-Hilbert plus Gibbons-Hawking terms) and use \eqref{BHFree} to get the free energy $F=(r_{+}+3r_{-})/4$. One then takes the derivative of $F$ with respect to $Q$ at fixed $T$, using \eqref{rpmRN} and \eqref{HawkTRN}. This non-trivially yields \eqref{muRN}.}

\subsection{\label{BBsubSec} The asymptotically flat black D3-brane geometry}

Let us now turn to another interesting example, the case of the asymptotically flat black D3-brane solution in type IIB supergravity.\footnote{We shall discuss the more general case of the $\alp$-corrected Schwarzschild-$\AdS_{5}$ geometry in Section \ref{alphaSec}.} The discussion is very similar to the case of the Reissner-Nordstr\"om black hole, and thus we shall be briefer.

The metric and Ramond-Ramond five-form field strength in the Euclidean are \cite{HS}
\begin{gather}
\label{metricBD3} \d s^{2} = \frac{f(\rho)\d t^{2} + \d\vec x^{2}}{\sqrt{H(\rho)}}+\sqrt{H(\rho)}\Bigl(\frac{\d\rho^{2}}{f(\rho)}+\rho^{2}\d\Omega_{5}^{2}\Bigr)\, ,\\
\label{F5BD3}
F_{5}=4iL^{4}\sqrt{1+\frac{\rho_{0}^{4}}{L^{4}}}\Biggl[\frac{\d\rho\wedge\d x^{1}\wedge\d x^{2}\wedge\d x^{3}\wedge\d t}{\rho^{5}H(\rho)^{2}}+i\omega_{\text{S}^{5}}\Biggr]\, ,
\end{gather}
where
\be
\label{fHD3B}H(\rho) = 1+\frac{L^{4}}{\rho^{4}}\, \cvp\quad
f(\rho) = 1-\frac{\rho_{0}^{4}}{\rho^{4}}\, \cdotp\ee
The metrics $\d\vec x^{2}$ and $\omega_{\text{S}^{5}}$ are the standard flat metric on $\mathbb R^{3}$ and volume form on $\text{S}^{5}$ respectively. The Euclidean time $t$ is periodic with period $\beta=1/T$, where
\be\label{TD3B} \beta=\frac{1}{T}=\pi\rho_{0}\sqrt{1+\frac{L^{4}}{\rho_{0}^{4}}} \, \cdotp\ee
The full geometry is spanned when $\rho_{0}\leq\rho<\infty$. The total charge of the solution corresponds to the number $N$ of D3 branes sourcing the geometry and is given by the standard formula
\be\label{ND3B} N=\frac{-i}{16\pi G_{10}}\frac{1}{\tau_{3}}\int_{\text{S}^{5}} \!\star F_{5}=
\frac{\pi L^{4}}{\ls^{4}\gs}\sqrt{1+\frac{\rho_{0}^{4}}{L^{4}}}\, \cvp\ee
where $G_{10}$, $\tau_{3}$, $\ls$, $\gs$ are the ten-dimensional Newton constant, D3-brane tension, string length and string coupling respectively, with the usual relations
\begin{align}\label{G10form} G_{10}&=\frac{\pi^{2}}{2}\gs^{2}\ls^{8}\, ,\\
\label{tau3}\tau_{3}&=\frac{1}{2\pi\ls^{4}\gs}\, \cdotp
\end{align}

Let us consider a BPS D3-brane probing the above background. Its tension is given by \eqref{tau3}. Its Euclidean action is
\be\label{probeD3B}\Sp = \tau_{3} A - i\tau_{3}\int\! C_{4}\, ,\ee
where the integral is taken along an arbitrary worldvolume spanned by $(t,x^{1},x^{2},x^{3})$, $C_{4}$ is a Ramond-Ramond potential satisfying $\d C_{4}=F_{5}$ and $A$ is the area of the worldvolume.\footnote{We orient the brane worldvolume as $(t,x^{1},x^{2},x^{3})$. We could also use a non-BPS probe with action $\hat\tau A - i\tau\int C_{4}$ and BPS bound $\hat\tau\geq |\tau|$ without changing the subsequent discussion in any important way.}

As in the case of the charged particle in the Reissner-Nordstr\"om geometry, the action of the probe brane is defined modulo the addition of an arbitrary worldvolume-independent constant. This ambiguity can be understood by noting that all the gauge potentials in the family
\be\label{C4D3B} C_{4}=i\sqrt{1+\frac{\rho_{0}^{4}}{L^{4}}}\biggl(\frac{1}{H(\rho)} + c\biggr)\d x^{1}\wedge\d x^{2}\wedge\d x^{3}\wedge\d t + \cdots\ee
yield the correct field strength \eqref{F5BD3}, for any value of the dimensionless constant $c$ (the $\cdots$ represent terms that do not contribute to the probe action). The undetermined constant is fixed along the lines of what we have done for the charged particle in the Reissner-Nordstr\"om geometry. We look at a brane sitting at a given value of $\rho$ in the asymptotically flat region $\rho\rightarrow\infty$ and impose that its energy per unit spatial volume reduces to its tension $\tau_{3}$ in this limit. This yields
\be\label{cBD3} c = -\lim_{\rho\rightarrow\infty}\frac{1}{H(\rho)} = -1\, .\ee
In particular, $C_{4}$ is singular at the horizon $\rho=\rho_{0}$. It is then straightforward to show that the minimum of the action is obtained for a schrunken brane sitting at $\rho=\rho_{0}$, yielding
\be\label{onshellBD3} \Sp^{*} = -\beta\tau_{3}V_{3}\sqrt{1+\frac{\rho_{0}^{4}}{L^{4}}}\biggl(\frac{1}{H(\rho_{0})} -1\biggr)=\frac{\beta\tau_{3}V_{3}}{\sqrt{1+\rho_{0}^{4}/L^{4}}}\,\cvp\ee
where $V_{3}=\int\!\d^{3}\vec x$.\footnote{Of course, this volume is strictly infinite. One could work instead with the action per unit volume, etc.} Comparing with \eqref{Spfact} and \eqref{murel}, with $q=1$ in our case,\footnote{It is natural to normalize the Ramond-Ramond charge in such a way that it simply counts the number of D3 branes.} we get the chemical potential
\be\label{muBD3} \mu = \frac{\tau_{3}V_{3}}{\sqrt{1+\rho_{0}^{4}/L^{4}}}\,\cdotp\ee
This is the correct known value for the solution \eqref{metricBD3}, \eqref{F5BD3}.

\section{\label{Sec3} Brane probes in holographic set-ups}

We now turn to cases involving probe branes in asymptotically AdS backgrounds, following \cite{FerRov}. Our goal is threefold: i) For pedagogical purposes and completeness, repeat the main arguments already presented in \cite{FerRov}; ii) Explain in great details the correct prescription given in \cite{FerRov} to fix the ambiguity in the probe actions in asymptotically AdS spaces. An important point will be to illustrate, on the example of the black D3-brane, how the AdS prescription actually follows from the flat space prescription after taking the near horizon limit; iii) Check the consistency of our fundamental relation for the D1/D5 system, M2 branes and M5 branes, which was not done explicitly in \cite{FerRov}.

\subsection{\label{sec3cons} Consequences of the fundamental relation}

Let us start with the standard case of the $\nn=4$ super Yang-Mills theory in four dimensions, with gauge group $\uN$,\footnote{Most of what we are going to say can actually be applied to any $\uN$ gauge theory \cite{fer1}.} which describes the low energy (or near horizon) dynamics of a stack of $N$ D3-branes. The generating functional of correlation functions (or the partition function) $Z=e^{-\Sg^{*}}$ has the standard 't~Hooft large $N$ expansion
\be\label{Zexpgauge} \ln Z = -\sum_{h\geq 0}N^{2-2h}F_{h}(\la)\, ,\ee
where $\la$ is the 't~Hooft coupling. For our purposes, the number of colors $N$ is identified with the total charge $Q$ (also counting the number of branes), and the thermodynamic limit corresponds to $N\rightarrow\infty$. Let us also take the $\la\rightarrow\infty$ limit, for which the usual gravitational bulk description is valid, and let us note $\lim_{\la\rightarrow\infty}F_{0}(\la) = F_{0}$. In these limits, \eqref{Zexpgauge} greatly simplifies to
\be\label{Zlimfor1} \Sg^{*}=-\ln Z = N^{2}F_{0}\, ,\ee
where $F_{0}$ does not depend on $N$. More generally, for other kinds of brane systems discussed in Section \ref{sec3Mbranes} below, the $N$-dependence can take the slightly more general form
\be\label{Zlimfor} \Sg^{*}=-\ln Z = N^{\gamma}F_{0}\, ,\ee
with some positive exponent $\gamma$. 

For all these cases, our fundamental relation \eqref{fundrel} greatly simplifies \cite{fer1}, because $\partial\Sg^{*}/\partial Q=\partial\Sg^{*}/\partial N$ is directly proportional to $\Sg^{*}$ itself. For one probe brane, we get
\be\label{frelsimple} \Sp^{*} = \frac{\gamma}{N}\Sg^{*}\, .\ee
This result is startling: the on-shell gravitational action and the on-shell probe action must be directly proportional, with a coefficient of proportionality which is fixed in terms of the scaling exponent governing the large $N$ behaviour of the free energy. For example, $\gamma=2$ for D3-branes.\footnote{Other interesting consequences of the fact that the free energy scales with $N^{2}$ has been recently discussed in \cite{karch}.}

\noindent\emph{Remark}: if we consider a probe anti-brane instead of a probe brane,  Eq.\ \eqref{frelsimple} is replaced by $\Sp^{*} = -\frac{\gamma}{N}\Sg^{*}$.

\subsection{\label{sec3BlackD3} The asymptotically AdS black D3-brane geometry}

\subsubsection*{The set-up}

Let us consider the geometry dual to the four dimensional $\nn=4$ gauge theory on flat space $\mathbb R^{3}$ at finite temperature $T$.\footnote{Note that this example is a limiting case of the AdS-Schwarzschild geometry studied in \cite{FerRov}.} It is obtained by taking the near-horizon limit of the geometry discussed in Section \ref{BBsubSec}. Formally, this near-horizon limit amounts to letting $L\rightarrow\infty$ in \eqref{metricBD3} and \eqref{F5BD3}, which yields
\begin{gather}
\label{metricBAdS} \d s^{2} = \frac{\rho^{2}}{L^{2}}\Bigl(f(\rho)\d t^{2} + \d\vec x^{2}\Bigr)+\frac{L^{2}}{\rho^{2}}\frac{\d\rho^{2}}{f(\rho)}+L^{2}\d\Omega_{5}^{2}\, ,\\
\label{F5BAdS}
F_{5}=\frac{4i}{L^{4}}\Bigl(\rho^{3}\d\rho\wedge\d x^{1}\wedge\d x^{2}\wedge\d x^{3}\wedge\d t+iL^{8}\omega_{\text{S}^{5}}\Bigr)\, .
\end{gather}
The full geometry, which is spanned when $\rho_{0}\leq\rho<\infty$, is of the form $M\times\text{S}^{5}$, the cigar-shaped bulk manifold $M=\text{B}^{2}\times\mathbb R^{3}$ being asymptotically $\text{AdS}_{5}$, with a boundary $X=\text{S}^{1}\times\mathbb R^{3}$. It is useful to note that
\be\label{F5better} F_{5}= \frac{4i}{L}\Omega_{M}-4L^{4}\omega_{\text{S}^{5}}\, ,\ee
where $\Omega_{M}$ is the volume form on $M$. The temperature and charge of the solution are given by
\be\label{TNBAdS} T=\frac{\rho_{0}}{\pi L^{2}}\,\cvp\quad N=\frac{\pi L^{4}}{\ls^{4}\gs}\,\cdotp\ee

\subsubsection*{Fixing the ambiguity in the action from the near-horizon limit}

As in Section \ref{BBsubSec}, the probe action is given by \eqref{probeD3B} and depends on a choice of gauge potential $C_{4}$. All the potentials of the form
\be\label{C4BAdS} C_{4} = i\Bigl(\frac{\rho^{4}}{L^{4}} + \tilde c\Bigr) \d x^{1}\wedge\d x^{2}\wedge\d x^{3}\wedge\d t + \cdots\ee
satisfy $\d C_{4}=F_{5}$ but yield different values for the action, parameterized by the dimensionless constant $\tilde c$. One thus faces again the problem of finding a prescription to fix this ambiguity. From the discussion of Sections \ref{RNsubSec} and \ref{BBsubSec}, it is clear that imposing smoothness of $C_{4}$ at $\rho=\rho_{0}$ does not make sense. Instead, one must use a condition at asymptotic infinity $\rho\rightarrow\infty$. 

Because the field strength does not vanish in this limit, the condition to be imposed in an asymptotically AdS space may not seem as obvious as in the case of an asymptotically flat space. A pedagogical way to guess the correct prescription is to use the following strategy: first find the correct result by directly taking the near horizon limit of the correct asymptotically flat space solution; then analyse the result and interpret it directly in asymptotically AdS space. 

We thus start from \eqref{C4D3B} with $c$ given by \eqref{cBD3} and let $L\rightarrow\infty$. Since the field strength \eqref{F5BAdS} is of order $1/L^{4}$, we keep all terms in $C_{4}$ of order $1/L^{4}$ or larger. This yields
\be\label{ctildecorrect} \tilde c = -\frac{\rho_{0}^{4}}{2L^{4}}-1\, .\ee
Plugging this result into the action for a brane sitting at a fixed value of $\rho$,\footnote{It is easy to check that the minimum of the action over all such worldvolumes coincides with the minimum of the action over all worldvolumes spanned by $t,x^{1},x^{2},x^{3}$.} we obtain
\be\label{pacBAdS} \Sp = \tau_{3}\beta V_{3}\biggl[\frac{\rho^{4}}{L^{4}}\Bigl(\sqrt{f(\rho)} - 1\Bigr) + \frac{\rho_{0}^{4}}{2 L^{4}} + 1\biggr]\, .\ee
Let us examine this result when $\rho\rightarrow\infty$. First, the terms proportional to $\rho^{4}$ cancel. This is simply the usual BPS condition. More interestingly, the $\rho_{0}$-dependent constant term is also canceled, due to the particular $\rho_{0}$-dependence in \eqref{ctildecorrect}. There remains a constant term, equal to $\tau_{3}\beta V_{3}$, whose temperature dependence comes entirely from the overall space-time volume $\beta V_{3}$ of the brane. This is a very special term: a so-called counterterm, that can be canceled by adding a cosmological constant to the brane action near the boundary.\footnote{We shall describe precisely the general form of these counterterms in Section \ref{sec3isop} below.} 

One is thus naturally led to the following condition to fix the ambiguity in the action: \emph{impose that the probe action reduces to a counterterm action near the AdS boundary.} The mathematically precise condition will be stated in the next subsection. Let us note that counterterms play a crucial role in the standard holographic dictionary, since they are required to make the on-shell gravitational action finite \cite{holren}. In view of the relation \eqref{frelsimple} between the on-shell gravitational action and the on-shell probe action that we want to obtain, it is satisfactory to find that counterterms do play a role in the precise definition of the probe action too.

These important points being understood, let us compute the minimal value of \eqref{pacBAdS}. It is easy to check that it is obtained for $\rho=\rho_{0}$. Using \eqref{tau3} and \eqref{TNBAdS}, which imply in particular that
\be\label{tau3AdS5} \tau_{3} = \frac{N}{2\pi^{2}L^{4}}\,\cvp\ee
we find
\be\label{SpsBAdS} \Sp^{*} = -\frac{N}{4}\pi^{2}\beta V_{3} T^{4} + \frac{N}{2\pi^{2}L^{4}}\beta V_{3}\, .\ee
Our fundamental relation \eqref{frelsimple}, with the correct value $\gamma=2$ for D3-brane (corresponding to $\ln Z\sim N^{2}$ at large $N$), together with $\Sg^{*} = \beta F$, finally yields the free energy
\be\label{freeBAdS} F = -\frac{N^{2}}{8}\pi^{2} V_{3} T^{4} + \frac{N^{2}}{4\pi^{2}L^{4}} V_{3}\, .\ee
\emph{The first term in the above equation matches precisely with the correct and well-known free energy of the planar $\nn=4$ super Yang-Mills theory at large 't~Hooft's coupling, obtained by computing $\Sg^{*}$ by the standard methods.} The second term corresponds to the contribution of a cosmological constant in the super Yang-Mills theory. This term can always be canceled by adding a local counterterm to the action and thus has no physical meaning.\footnote{In particular the precise numerical coefficient $N^{2}/(4\pi^{2})$ that we have found above by looking at the near horizon limit of the asymptotically flat geometry does not have any physical meaning in the asymptotically AdS set-up. It could be set to any number we wish.}

\subsection{\label{sec3isop} General asymptotically AdS spaces and the isoperimetric inequality}

Following \cite{FerRov}, we are now going to greatly generalize the above discussion and show that the fundamental relation \eqref{frelsimple} is consistent in any relevant Einstein-Poincar\'e space.\footnote{This is the so-called ``pure gravity'' case. See Section \ref{D3dilsec} for a generalization including a non-trivial dilaton and Section \ref{CSec} for a brief discussion of possible extensions of the isoperimetric inequality used below for more general supergravity backgrounds.} We shall see that consistency is made possible by an interesting geometric property of these spaces, a non-trivial isoperimetric inequality first derived in \cite{Wang}.

\subsubsection*{The set-up}

\begin{figure}
\centerline{\includegraphics[width=6in]{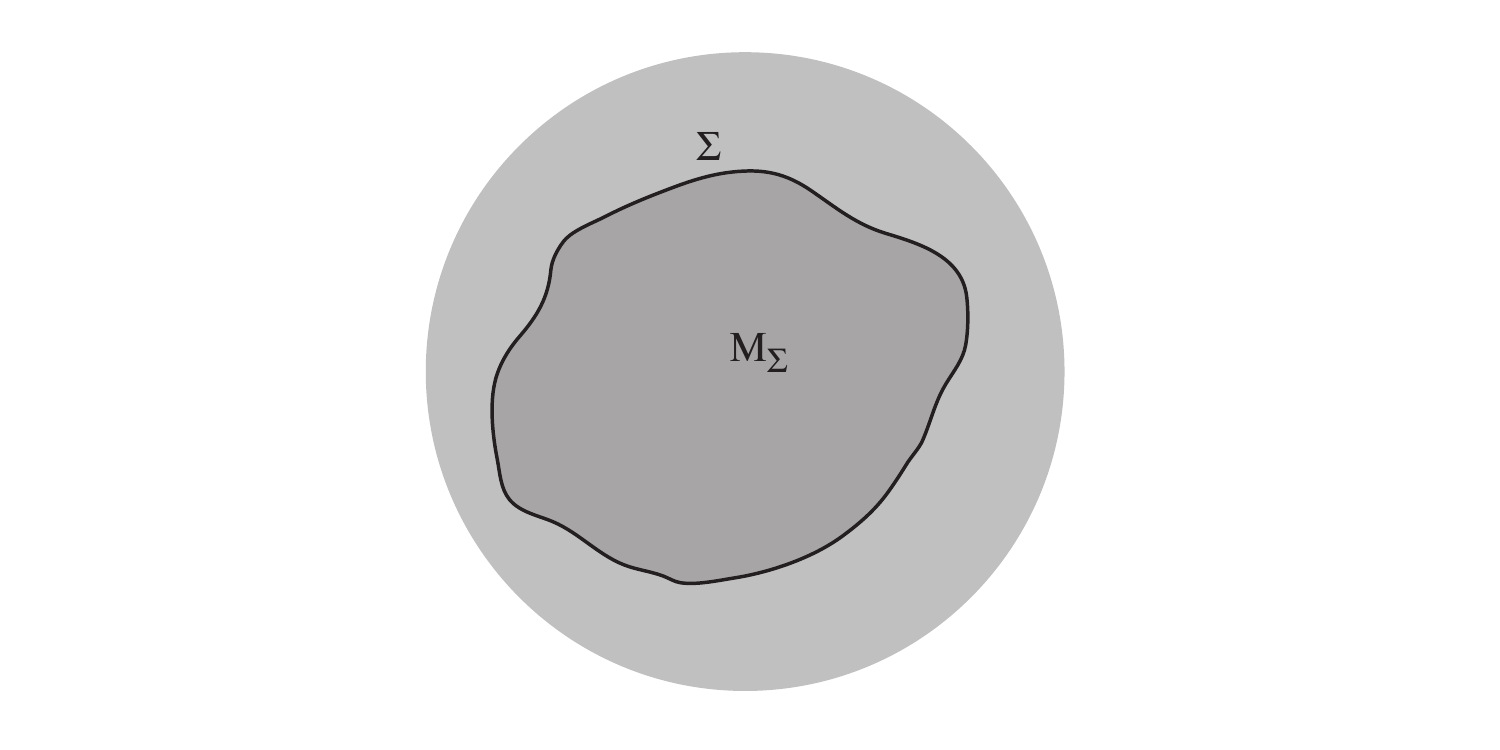}}
\caption{An arbitrary $(d-1)$-brane embedded into the bulk asymptotically AdS space. The brane worldvolume $\Sigma$ encloses the volume $V(M_{\Sigma})$ of bulk space, depicted in dark grey.
\label{figbrane}}
\end{figure}

We consider a general asymptotically AdS Euclidean bulk space $M$ of dimension $d+1$, with $d$-dimensional boundary $X=\partial M$ endowed with a conformal class of metrics $[\bar g]$.\footnote{We choose the boundary $X$ to be compact. Non-compact boundaries, as in the example $X=\text{S}^{1}\times\mathbb R^{3}$ of Section \ref{sec3BlackD3}, can be obtained by taking the large volume limit of compact boundaries.}  The bulk metric $G$ satisfies the Einstein-Poincar\'e condition
\be\label{PEcond} R_{\mu\nu} = -\frac{d}{L^{2}} G_{\mu\nu}\, ,\ee
where $L$ is the scale of the asymptotic AdS space. The conformal class $[\bar g]$ on $X$ is chosen to have a non-negative Yamabe invariant,\footnote{The non-negativity of the Yamabe invariant is equivalent to the fact that the action for a conformally coupled scalar on the boundary is bounded from below \cite{WY}. This condition is required for the stability of the boundary CFT. Mathematically, it is equivalent to the following fact. By the Trudinger-Aubin-Schoen theorem, there always exists a representative of the conformal class on the boundary having constant scalar curvature. This scalar curvature must be non-negative.} but is otherwise arbitrary. We consider a $(d-1)$-brane probing the geometry, whose worldvolume $\Sigma\subset M$ is homologous to the boundary $X$, see Fig.\ \ref{figbrane}. Its Euclidean action is given by
\be\label{probegenact} \Sp = \tau_{d-1} A(\Sigma) - i\tau_{d-1}\int_{\Sigma}\! C_{d}\, ,\ee
where $A(\Sigma)$ is the volume of the worldvolume for the induced metric on $\Sigma$ and $C_{d}$ is a gauge potential to which the brane couples.\footnote{If worldvolume gauge fields are included, $A(\Sigma)$ must be replaced by the more general Dirac-Born-Infeld action. It is straightforward to show that this more general action is always greater than or equal to $A(\Sigma)$. Since we shall be interested in the on-shell, minimum value of the action only, this makes worldvolume gauge fields irrelevant for our purposes.}$^{,}$\footnote{We can also use a non-BPS probe with action $\hat\tau A - i\tau\int C_{d}$ and BPS bound $\hat\tau\geq |\tau|$ without changing the subsequent discussion in any important way.} The gauge potential satisfies
\be\label{dCd} \d C_{d}= F_{d+1}= i\frac{d}{L}\Omega_{M}\, ,\ee
where $\Omega_{M}$ is the volume form of $M$, generalizing \eqref{F5better}.\footnote{The second term in \eqref{F5better} is irrelevant for our purposes.} 

The ambiguity in $C_{d}$, coming from the integration of \eqref{dCd}, produces the usual ambiguity in the probe action. Since all brane worldvolumes we consider are homologous to each other, this ambiguity is simply an overall worldvolume-independent constant in the action. Up to this constant, that we denote by $s$, \eqref{dCd} and Stokes' theorem imply that 
\be\label{SpSt} \Sp(\Sigma) = \tau_{d-1}\Bigl(A(\Sigma) - \frac{d}{L}V(M_{\Sigma})\Bigr) + s\, ,\ee
where $V(M_{\Sigma})$ is the volume of bulk space enclosed by $\Sigma$, as depicted in Fig.\ \ref{figbrane}.

\subsubsection*{Fixing the ambiguity in the action}

The general prescription to fix the ambiguity in the probe action in an arbitrary asymptotically AdS space follows from the discussion in Section \ref{sec3BlackD3}: \emph{we impose that the probe action goes to a purely counterterm action near the boundary of AdS.}

The precise implementation of this prescription is as follows. First, one uses Fefferman-Graham coordinates near the boundary. In these coordinates, the bulk metric can be written
\be\label{PEmet} G = \frac{L^{2}\d r^{2}+ g}{r^{2}}\, \cvp\ee
where
\be\label{FGexp} g(r,z)=\bar g(z) + g_{(2)}(z) r^{2} + \cdots\ee
can be expanded near the boundary at $r=0$ and we denote by $z$ the coordinates on the boundary. Next, we consider a brane worldvolume $\Sigma_{\epsilon}$ given by $r=\epsilon$ and denote by $g_{\epsilon}$ the induced metric on $\Sigma_{\epsilon}$. A \emph{counterterm action} is an action of the form
\be\label{Sctdef} \Sct(\Sigma_{\epsilon}) = \int_{\Sigma_{\epsilon}}\!\d^{d}z\sqrt{\det g_{\epsilon}}\Bigl(c_{d}L^{-d} + c_{d-2}L^{-d+2}R[g_{\epsilon}] + \cdots\Bigr)\, ,\ee
for which the coefficients $c_{d}$, $c_{d-2}$, etc., are dimensionless numbers that may depend on $\epsilon$ but only on $\epsilon$ and may at worst diverge logarithmically when $\epsilon\rightarrow 0$ (the power-like divergences come from the factor $\sqrt{\det g_{\epsilon}}$ in \eqref{Sctdef} and their general form is consistent with the standard power-counting arguments). We have denoted by $R[g_{\epsilon}]$ the scalar curvature constructed from the metric $g_{\epsilon}$. The $\cdots$ represent similar but higher derivative local curvature terms built from the metric $g_{\epsilon}$, of dimension less than $d$. Our prescription to fix the constant $s$ in \eqref{SpSt} is then to impose that
\be\label{genpres} \lim_{\epsilon\rightarrow 0}\bigl(\Sp(\Sigma_{\epsilon})-\Sct(\Sigma_{\epsilon})\bigr) = 0\, , \ee
for some counterterm action $\Sct$.

\noindent\emph{Example}: for pedagogical purposes, let us revisit the example of Section \ref{sec3BlackD3}. The radial Fefferman-Graham coordinate $r$ is related to the coordinate $\rho$ used in \eqref{metricBAdS} and \eqref{F5BAdS} by
\be\label{rrhoeq} r^{2}=\frac{2L^{2}}{\rho^{2}+\sqrt{\rho^{4}-\rho_{0}^{4}}}\, \cdotp\ee
This yields the five-dimensional bulk metric
\be\label{FGmet} \d s^{2}= \frac{1}{r^{2}}\Biggl[L^{2}\d r^{2} + \frac{\Bigl(1-\frac{\rho_{0}^{4}r^{4}}{4L^{4}}\Bigr)^{2}}{1+\frac{\rho_{0}^{4}r^{4}}{4L^{4}}}\d t^{2} + \Bigl(1+\frac{\rho_{0}^{4}r^{4}}{4L^{4}}\Bigr)\d\vec x^{2}\Biggr]\, .\ee
By evaluating \eqref{SpSt} for a brane at fixed $r$, using \eqref{TNBAdS} and \eqref{tau3AdS5}, we get
\be\label{Sprex3} \Sp = \frac{N}{4}\pi^{2}\beta V_{3}T^{4}\Bigl(1-\frac{1}{4}\pi^{4}L^{4}T^{4}r^{4}\Bigr) + s\, .\ee
In particular, 
\be\label{Sprexeps} \Sp(\Sigma_{\epsilon}) = \frac{N}{4}\pi^{2}\beta V_{3}T^{4} + s + O(\epsilon^{4})\, .\ee
On the other hand, the induced metric on $\Sigma_{\epsilon}$, derived from \eqref{FGmet}, is flat. All its local curvature invariants thus vanish. The most general counterterm action is then a cosmological constant term, which takes the form
\be\label{Sctex3}\Sct = c_{4}(\epsilon)L^{-4}A(\Sigma_{\epsilon}) = \frac{\beta V_{3}}{L^{4}}\frac{c_{4}(\epsilon)}{\epsilon^{4}} + O(\epsilon^{4})\,.\ee
Note that a crucial point here is that $c_{4}$ can only depend on $\epsilon$, but not on other parameters like the temperature. Comparing \eqref{Sprexeps} and \eqref{Sctex3}, we see that our prescription \eqref{genpres} implies that
\be\label{svalueex1} s = - \frac{N}{4}\pi^{2}\beta V_{3}T^{4} + c\frac{\beta V_{3}}{L^{4}}\, \cvp\ee
for an arbitrary dimensionless numerical constant $c$. Moreover, \eqref{Sprex3} implies that $\Sp^{*}=s$, since the maximal value of $r^{4}$ defined by \eqref{rrhoeq} is $4/(\pi L T)^{4}$. This is perfectly consistent with the result of Section \ref{sec3BlackD3}.

\noindent\emph{The general case}: the above calculation can be straightforwardly generalized to an arbitrary asymptotically AdS geometry. First, let us note that, from dimensional analysis, the brane tension $\tau_{d-1}$ will always be of the form $c_{d}L^{-d}$ for some dimensionless numerical constant $c_{d}$. When evaluated on $\Sigma_{\epsilon}$, the area term $\tau_{d-1}A$ in the probe action is thus automatically a counterterm. The condition \eqref{genpres} thus simply yields
\be\label{sgenvalue} s = \lim_{\epsilon\rightarrow 0}\Bigl(\frac{d}{L}\tau_{d-1}V(M_{\Sigma_{\epsilon}}) + \Sct(\Sigma_{\epsilon})\Bigr)\, ,\ee
for some counterterm action $\Sct$. Let us note that, of course, $\lim_{\epsilon\rightarrow 0}V(M_{\Sigma_{\epsilon}})$ is infinite, but it is always possible to choose $\Sct(\Sigma_{\epsilon})$ to cancel the infinities, as implied by the standard holographic renormalization procedure \cite{holren}.

\subsubsection*{The isoperimetric inequality}

The last step is to compute the minimal value of the probe action $\Sp$. From \eqref{SpSt} and \eqref{sgenvalue}, we find that
\be\label{Spstargen} \Sp^{*} = \tau_{d-1}\min_{\Sigma}\Bigl(A(\Sigma) - \frac{d}{L}V(M_{\Sigma})\Bigr) +  \lim_{\epsilon\rightarrow 0}\Bigl(\frac{d}{L}\tau_{d-1}V(M_{\Sigma_{\epsilon}}) + \Sct(\Sigma_{\epsilon})\Bigr)\, ,\ee
where the quantity $A-\frac{d}{L}V$ must be minimized over all possible worldvolumes (i.e.\ embedded hypersurfaces homologous to the boundary) in the bulk space. At first sight, this minimization problem might seem intractable, but it turns out that the solution is actually very simple and elegant. Indeed, there exists a so-called isoperimetric inequality, stating that
\be\label{isopineq} A(\Sigma)\geq \frac{d}{L}V(M_{\Sigma})\ee
for any embedded hypersurface $\Sigma$ in any asymptotically AdS space with a non-negative Yamabe invariant on the boundary. This inequality was first proven in \cite{Wang}; a simpler proof was also provided in \cite{FerRov}. It immediately implies that the minimum of $A-\frac{d}{L}V$ is zero, a value that can always be realized by considering a schrunken brane. Overall, we have thus found that
\be\label{Spstarfin} \Sp^{*} = \lim_{\epsilon\rightarrow 0}\Bigl(\frac{d}{L}\tau_{d-1}V(M_{\Sigma_{\epsilon}}) + \Sct(\Sigma_{\epsilon})\Bigr)\, .\ee

\noindent\emph{Remark}: if we consider a probe anti-brane instead of a probe brane, the functional that we need to minimize is $A(\Sigma) + \frac{d}{L}V(M_{\Sigma})$, whose minimum is trivially zero. The discussion then proceeds along the line of the case of the probe brane, see also the remark at the end of Section \ref{sec3cons}.

\subsubsection*{Checking the fundamental identity}

To check our fundamental identity \eqref{frelsimple}, let us first recall how the on-shell gravitational action is computed. The Einstein-Poincar\'e condition \eqref{PEcond} follows from the usual Einstein-Hilbert action with a suitable cosmological constant term,
\be\label{SgEH1} \Sg = -\frac{1}{16\pi G_{d+1}}\int_{M}\!\d^{d+1}x\sqrt{\det G}\,\Bigl(R +\frac{d(d-1)}{L^{2}}\Bigr)\, .\ee
Computing $R$ from \eqref{PEcond} yields
\be\label{ShEH2} \Sg^{*} = \frac{d}{8\pi G_{d+1}L^{2}}V(M)\, ,\ee
where $V(M)$ is the volume of space-time. Of course, this volume is infinite. The correct definition of the on-shell gravitational action requires regularization and renormalization. The procedure is standard \cite{holren}. One introduces the Fefferman-Graham coordinates and replace the non-compact space-time $M$ by the compact region $M_{\epsilon}$, defined to be the set of points having $r\geq\epsilon$. This compact region is identical to the region $M_{\Sigma_{\epsilon}}$ considered previously. The infinities in $V(M_{\Sigma_{\epsilon}})$ when $\epsilon\rightarrow 0$ are then canceled by adding to the Einstein-Hilbert action a counterterm action of the form \eqref{Sctdef}.\footnote{Since the compact space $M_{\Sigma_{\epsilon}}$ has a boundary, one may note that in principle the Einstein-Hilbert term must be supplemented by a boundary Gibbons-Hawking term in the gravitational action. However, a direct calculation shows that this term always reduces to a pure counterterm when $\epsilon\rightarrow 0$. This is a nice property of the hypersurfaces $M_{\Sigma_{\epsilon}}$ which are defined by using the radial Fefferman-Graham coordinate.} This yields
\be\label{Sgstarfinal} \Sg^{*} =\lim_{\epsilon\rightarrow 0}\Bigl( \frac{d}{8\pi G_{d+1}L^{2}}V(M_{\Sigma_{\epsilon}}) + \Sct(\Sigma_{\epsilon})\Bigr)\, .\ee
Comparing with \eqref{Spstarfin}, we see that the fundamental relation \eqref{frelsimple} is satisfied if and only if \cite{FerRov}
\be\label{gammaform} \gamma = 8\pi N L G_{d+1}\tau_{d-1}\, .\ee
So everything boils down to checking a seemingly mysterious but very simple numerical relation between the bulk Newton constant $G_{d+1}$ and the brane charge $\tau_{d-1}$!

Before we go on to check \eqref{gammaform} explicitly in a variety of cases, let us emphasize how two important puzzles with the fundamental relation \eqref{frelsimple} have been solved by the above discussion. 

The first puzzle concerns the holographic renormalization procedure. It is essential in making the on-shell gravitational action finite and it also implies a certain ambiguity related to the possibility of adding finite local counterterms. If the on-shell gravitational action is to be identified with the on-shell probe action through \eqref{frelsimple}, this important feature must have a counterpart for the on-shell probe action. At first sight, this is rather mysterious. The resolution of the puzzle comes from realizing that the probe action does suffer from an ambiguity, coming from the choice of the gauge potential coupling to the brane. The way to fix this ambiguity in asymptotically AdS spaces is to impose that the brane action goes to a counterterm action near the boundary. This prescription actually fixes the ambiguity only partially, since it is always possible to add finite local counterterms. The result is that both $\Sg^{*}$ and $\Sp^{*}$ share the same properties relative to holographic renormalization. There is no obstruction in making an identification like \eqref{frelsimple}.

The second puzzle concerns the computation of the minimal value $\Sp^{*}$ of the brane action. Naively, this looks like a very complicated problem, strongly depending on the details of the bulk geometry. But this difficulty is surmounted thanks to the remarkable geometric property of AdS spaces coded in the isoperimetric inequality \eqref{isopineq}.

\subsection{\label{sec3Mbranes} Examples of D-brane and M-brane systems}

We have just shown that the fundamental relation \eqref{frelsimple} between the on-shell probe action $\Sp^{*}$ and the on-shell gravitational action $\Sg^{*}$ is automatically satisfied provided the algebraic constraint \eqref{gammaform} is valid. The values of the parameters involved in this relation are \emph{independent} of the details of the geometry and are thus given once and for all by the microscopic definition of the system. In this Section, we verify the relation \eqref{gammaform} for various D-brane and M-brane systems, providing further non-trivial checks of our general framework. Moreover, beyond verification purposes, they illustrate new interesting features that were not present in the explicit example of Section \ref{sec3BlackD3}.

The geometries we consider are of the form $\M\times K$, where $\M$ is asymptotically $\text{AdS}_{d+1}$ and $K$ is a compact manifold, on which for simplicity all the fields are assumed to be constant. The effective Newton constant $G_{d+1}$ on $\M$ is then given by
\be\label{newtonconst}
	G_{d+1} = \frac{\GN}{e^{-2\phi}\Vol(K)}\, \cvp
\ee
where $\phi=\Phi-\log \gs$ is the dilaton and $\GN$ is the Newton constant of the original, non-reduced theory. In our examples $\GN$ will be ten dimensional $G_{10}$ or eleven dimensional $G_{11}$ (in which case there is of course no dilaton, so we simply set $\phi=0$ in \eqref{newtonconst}) gravitational constant. Explicit expressions in our conventions can be found in Appendix \ref{convApp}. 

Our general analysis of Section \ref{sec3isop} remains valid for constant dilaton. The only difference is an additional factor of $e^{-\phi}$ in the DBI term in \eqref{probegenact}, which now reads
\be\label{DBIdil}
	\tau_{d-1} e^{-\phi} A(\Sigma)\, .
\ee
As a consequence, \eqref{gammaform} becomes
\be\label{gammaformdil} \gamma = 8\pi N L G_{d+1}e^{-\phi}\tau_{d-1}\, .\ee
Of course, for non-constant dilaton, the analysis must be adapted. A simple example with non-constant dilaton will be considered in Section \ref{D3dilsec}.

\subsubsection*{D3 branes}

In this case, recall from \eqref{TNBAdS} that the charge is
\be\label{char00} N = \frac{-i}{16\pi G_{10}}\frac{1}{\tau_{3}}\int_{\text{S}^{5}}\!\star F_{5} =2\pi^{2}L^{4}\tau_{3} = \frac{\pi L^{4}}{\gs\ls^{4}}\,\cvp\ee
whereas the five-dimensional Newton constant is given by
\be\label{Gnewton5}
	G_{5} = \frac{G_{10}}{L^{5}\Vol(\text{S}^{5})} = \frac{\pi L^{3}}{2 N^{2}}\,\cdotp
\ee
To check \eqref{gammaformdil}, we thus compute
\be\label{D3prex} 8\pi N L G_{5}\tau_{3} = 8\pi N L \times \frac{\pi L^{3}}{2 N^{2}}\times\frac{N}{2\pi^{2}L^{4}} = 2\, ,\ee
consistently with the correct value $\gamma=2$ for the D3 branes. This is of course in line with the special case studied in Section \ref{BBsubSec}.

\subsubsection*{D1 and D5 branes}

We consider the standard near-horizon D1/D5 geometry \cite{malda}, of the form $\M \times \text{T}^{4} \times \text{S}^{3}$, where $\M$ is asymptotically $\AdS_{3}$ of radius $L$ and $S^{3}$ is the three-sphere of radius $L$. In this case, we have two charges, associated with the numbers $N_{1}$ and $N_{5}$ of D1 and D5 branes. Thus we have two versions of the relation \eqref{gammaformdil} that we can check, according to which type of probe brane we use. This also means that the on-shell gravitational action $\Sg^{*}$ can be obtained from \eqref{frelsimple} using either D1 or D5 probes.

On the one hand, let us note that, as is well-known, the free energy scales as $N_{1}N_{5}$ and thus the exponents entering \eqref{gammaformdil} are
\be\label{gammaD1D5}
	\gamma_{\text{D1}}=\gamma_{\text{D5}}=1\, .
\ee
On the other hand, the Newton constant and dilaton are determined in terms of the compact factor of the geometry alone. For example, we can consider the vacuum $\text{AdS}_{3}\times\text{T}^{4}\times\text{S}^{3}$ geometry, given by
\begin{subequations}\label{decgeom}
\begin{align}
\label{decgeommet}
	\d s^{2} & = \frac{L^2}{r^{2}} \d r ^{2}+ \frac{r^{2}}{L^2} ( \d t^{2} +\d x^{2}) + e^\phi \d z^{a} \d z^{a} + L^2 \d \Omega_{3}^{2}\, , \\
	F_{3} &= \frac{2L^{2}}{ e^{\phi}} \big( i\omega_{\AdS_{3}} + \omega_{\text{S}^{3}} \big)\, .
\end{align}
\end{subequations}
We denote by $\omega_{\AdS_{3}}$ and $\omega_{\text{S}^{3}}$ the volume forms on the spaces $\AdS_{3}$ and $\text{S}^{3}$ of unit radii. The dilaton fluctuation $\phi$ is constant. The torus coordinates $z^{a}$ are periodic,
\be\label{T4}
	z^{a} \sim z^{a} + 2\pi \rho\,,
\ee
where $\rho$ is an arbitrary length scale. The charges are given by\footnote{Compare with \eqref{ND3B}.}
\begin{align}
 \label{chargeE}	N_{1} &= \frac{-i}{16\pi G_{10}}\frac{1}{\tau_{1}}\int_{\text{S}^{7}} \!\star F_{3} = 2\pi L^{2}\tau_{5}\Vol(\text{T}^{4}) e^{-\phi}=
 \frac{L^{2}}{2\pi\gs\ls^{6}}(2\pi\rho)^{4}e^{\phi} \, , \\
 \label{chargeM}	N_{5} &=  \frac{-i}{16\pi G_{10}}\frac{1}{\tau_{5}}\int_{\text{S}^{3}} \!\star F_{7} = \frac{1}{16\pi G_{10}}\frac{1}{\tau_{5}}\int_{\text{S}^{3}}\! F_{3} =2\pi L^{2}\tau_{1}e^{-\phi}=
 \frac{2\pi L^{2}}{\gs\ls^{2}}e^{-\phi}\, ,
\end{align}
where we have used \eqref{G10form} and the values of the brane tensions
\be\label{tensionD1D5}
	\tau_{1} = \frac{1}{\ls^{2}\gs} \, \cvp\quad \tau_{5} = \frac{1}{(2\pi)^{2}\ls^{6}\gs}\, \cdotp
\ee
The three-dimensional Newton constant is then
\be\label{G3}
	G_{3} = \frac{G_{10}}{e^{-2\phi}L^{3} \Vol(\text{T}^{4}\times \text{S}^{3})} = \frac{L}{4N_{1}N_{5}}\, \cdotp
\ee

We can now check \eqref{gammaformdil}. For a D1-brane probe, we find, using in particular \eqref{chargeM} and \eqref{tensionD1D5},
\be\label{D1probe}
	8\pi N_{1}L G_{3} e^{-\phi}\tau_{1} = 8\pi N_{1} L\times \frac{L}{4N_{1}N_{5}}\times\frac{N_{5}}{2\pi L^{2}}= 1\, ,
\ee
matching perfectly the value of $\gamma$ for the D1-brane, see \eqref{gammaD1D5}. For a D5-brane probe, the tension $\tau_{1}$ appearing in \eqref{gammaformdil} is of course the effective tension $\tau_{5}\Vol(\text{T}^{4})$ of the D5 wrapped on $\text T^{4}$. Using \eqref{chargeE} and \eqref{tensionD1D5}, this yields
\be\label{D5probe} 8\pi N_{5}LG_{3}e^{-\phi}\bigl(\tau_{5}\Vol(\text{T}^{4})\bigr) = 8\pi N_{5}L\times \frac{L}{4N_{1}N_{5}}\times \frac{N_{1}}{2\pi L^{2}} = 1\, ,
\ee
again matching perfectly the value of $\gamma$ for the D5-brane given by \eqref{gammaD1D5}.

\subsubsection*{M2 branes}

The near horizon M2-brane geometry \cite{malda} is $M\times\text{S}^{7}$ where $M$ is asymptotically $\AdS_{4}$ of radius $L$ and $\text{S}^{7}$ is the seven-sphere of radius $2L$. For example, the vacuum solution is 
\be\label{M2geom}
	\d s^{2} = \frac{L^{2}}{r^{2}}\d r^{2} + \frac{r^{2}}{L^{2}}\bigl(\d t^{2} + \d\vec x^{2}\bigr) + 4 L^{2} \d \Omega_{7}^{2}\, , \quad 	F_{4} = \frac{3i}{L} \omega_{\AdS_{4}} \, . \ee
The charge is computed as
\be\label{chM2} N = \frac{-i}{16\pi G_{11}}\frac{1}{\tau_{M2}}\int_{\text{S}^{7}}\!\star F_{4} = 2^{15/3}\pi^{2} L^{6}\tau_{M2}^{2} = \frac{2^{11/3}\pi^{8/3}L^{6}}{\ell_{11}^{6}}\, \cvp\ee
%
where we have used the standard formula for the M2-brane tension in terms of the eleven-dimensional Planck length $\ell_{11}=G_{11}^{1/9}$,
\be\label{Meten} \tau_{M2} = \frac{\pi^{1/3}}{2^{2/3}\ell_{11}^{3}}\,\cdotp\ee
Using \eqref{chM2}, the four-dimensional Newton constant is
\be\label{G4for} G_{4} = \frac{\ell_{11}^{9}}{(2L)^{7}\Vol(\text{S}^{7})} = \frac{3 L^{2}}{2^{3/2}N^{3/2}}\,\cdotp\ee
%
Using again \eqref{chM2}, Eq.\ \eqref{gammaformdil} thus yields
\be\label{checkM2} 8\pi N L G_{4} \tau_{M2} = 8\pi N L \times \frac{3 L^{2}}{2^{3/2}N^{3/2}} \times \frac{N^{1/2}}{2^{15/6}\pi L^{3}} = \frac{3}{2}\,\cdotp\ee
%
Remarkably, this is consistent with the well-known $N^{3/2}$ scaling of the free energy for the M2 branes.

\subsubsection*{M5 branes}

The near horizon M5-brane geometry \cite{malda} is $M\times \text{S}^{4}$ where $M$ is asymptotically AdS$_{7}$ of radius $L$ and S$^{4}$ is the four-sphere of radius $L/2$. For example, the vacuum solution is
\be\label{M5geom}
	\d s^{2} = \frac{L^{2}}{r^{2}} \d r^{2} + \frac{r^{2}}{L^{2}} \Big( \d t^{2}+ \d \vec x ^{2} \Big) + \frac{L^{2}}{4} \d \Omega_{4}^{2}\, , 	\quad F_{7}  = \frac{6i}{L} \omega_{\AdS_{7}}\, .
\ee
The charge is computed as
\be\label{chM5}
	N= \frac{-i}{16 \pi G_{11}} \frac{1}{\tau_{M5}} \int_{\text{S}^{4}} \star F_{7} = \sqrt{\frac{\pi^{3}\tau_{M5}}{2}} L^{3} = \frac{\pi^{4/3}}{2^{5/3}} \frac{L^{3}}{\ell_{11}^{3}} \, \cvp
\ee
where we have used the standard formula for the M5-brane tension in terms of the eleven dimensional Planck length $\ell_{11}=G_{11}^{1/9}$,
\be\label{M5planck}
	\tau_{M5} = \frac{1}{2^{7/3}\pi^{1/3}\ell_{11}^{6}}\,  \cdotp
\ee
Using \eqref{chM5}, the seven-dimensional Newton constant is
\be\label{GN7}
	G_{7} = \frac{\ell^{9}_{11}}{(L/2)^{4} \text{Vol}(\text{S}^{4})} = \frac{3 \pi^{2}L^{5}}{16 N^{3}}\, \cdotp
\ee
Using again \eqref{chM5}, Eq.\ \eqref{gammaformdil} thus yields
\be\label{checkM5}
	8 \pi N L G_{7} \tau_{M5} = 8 \pi N L \times \frac{3\pi^{2}L^{5}}{16 N^{3}} \times \frac{2 N^{2}}{\pi^{3}L^{6}} = 3\, .
\ee
This is in perfect agreement with the well-known $N^{3}$ scaling of the free energy for the M5 branes.

\subsection{A simple example with a non-trivial dilaton}\label{D3dilsec}

In this last subsection, we present a simple generalization for which we allow a non-constant dilaton $\phi$. We consider a ten-dimensional space-time of the form $M\times \text{S}^{5}$, where $M$ is asymptotically AdS$_{5}$ with radius $L$ and S$^{5}$ is the five-sphere of radius $L$. Assuming that the ten-dimensional supergravity fields remain constant on S$^{5}$, the action for the fields on $M$ (and using for convenience the Einstein frame metric $g$) reads
\be\label{Sfive}
	\Sfive = -\frac{1}{16 \pi G_{5}} \int_{M} \d^{5} x \sqrt{ g} \Big(  R( g) - 2 \Lambda - \frac{1}{2} |\d \phi|_{ g}^{2}  \Big)\, ,
\ee
where the cosmological constant is $\Lambda = -6/L^{2}$ and the five-dimensional Newton constant is given by
\be\label{G5}
	G_{5} = \frac{G_{10}}{\Vol(S^{5}_{L})} \, \cdotp
\ee
The equations of motion derived from \eqref{Sfive} reads
\be\label{eqM}
	R(g)_{\mu\nu} = \frac{1}{2} \del_{\mu} \phi\, \del_{\nu} \phi - \frac{4}{L^{2}}  g_{\mu\nu} \, , \quad \Delta_{ g} \phi = 0 \, .
\ee
The on-shell value of the gravitational action therefore has the same form as in \eqref{ShEH2}, namely
\be\label{Sfivestar}
	\Sfive^{*} = \frac{1}{2 \pi G_{5}L^{2}} V(M)\, .
\ee
On the other hand, the probe action for a D3 brane is as in \eqref{probegenact}, namely
\be\label{probeD3dil}
	\Sp = \tau_{3} A(\Sigma) - i \tau_{3} \int_{\Sigma} C_{4}\, ,
\ee
where the area $A(\Sigma)$ is computed using the metric induced on $\Sigma$ from the Einstein frame metric $g$ and $C_{4}$, as usual, is such that \eqref{F5better} holds.\footnote{Note that the dilaton does not appear explicitly in the D3-brane action in the Einstein frame.}

The discussion of Section \ref{sec3isop} can then be repeated straightforwardly. The important point is that the equations of motion \eqref{eqM} implies that
\be\label{isocond}
	R_{\mu\nu} + \frac{4}{L^{2}} g_{\mu\nu} = \frac{1}{2} \del_{\mu } \phi \, \del_{\nu} \phi \geq 0\, .
\ee
As explained in \cite{FerRov}, this condition ensures the validity of the isoperimetric inequality \eqref{isopineq} and thus, also using \eqref{D3prex}, of our fundamental relation \eqref{frelsimple}.

\section{\label{alphaSec} Schwarzschild-$\AdS_{5}$ with $\alp$-corrections}
\subsection{General consequences of the fundamental relation}

Until now, we have verified the general formula \eqref{fundrel} relating the on-shell probe action $\Sp^{*}$ to the on-shell supergravity action $\Sg^{*}$ in the regime where the supergravity approximation for the dual bulk description is reliable. In the holographic set-up coming from D3 branes in superstring theory, this corresponds to the strong 't~Hooft coupling regime $\la\rightarrow\infty$. Our fundamental relation is then equivalent to the simpler relation \eqref{frelsimple}.

The goal of the present section is to take into account the first non-trivial $\alpha'$ corrections to supergravity or, equivalently, the first non-trivial corrections to the $\la\rightarrow\infty$ limit, still staying in the thermodynamic limit $N\rightarrow\infty$. Eq.\ \eqref{Zlimfor} is then replaced by
\be\label{hol1} \Sg^{*} = N^{2}F_{0}(\la)\, ,\ee
where we keep explicitly the $\la$-dependence in $F_{0}$. The fundamental relation \eqref{fundrel} thus yields 
\be\label{hol2}\frac{\partial \bigl(N^{2}F_{0}(\la)\bigr)}{\partial N} = \Sp^{*}\, .\ee
At this stage, it is important to recall that the 't~Hooft coupling $\la$ depends itself on $N$ via the standard relation
\be\label{lagsrel} \la = 4\pi \gs N\ee
and thus \eqref{hol2} is equivalent to
\be\label{fundrelbis} \Sp^{*} =  N \bigl(2F_{0}(\la) +  \la  F'_{0}(\la)\bigr)\, .\ee
This can be conveniently rewritten as
\be\label{frc2} \frac{\partial}{\partial\la}\bigl(\la^{2} \Sg^{*}\bigr)= N\la\Sp^{*} \, .\ee
Equations \eqref{fundrelbis} and \eqref{frc2} are highly non-trivial predictions in the theory at finite $\la$ or, equivalently, at finite $\alpha'$. In particular, the actions $\Sp^{*}$ and $\Sg^{*}$ entering these equations are the $\alp$-corrected D-brane action and supergravity action, evaluated on the $\alp$-corrected supergravity background. 


\noindent\emph{Remarks}:

\noindent\ i) At large $\la$, we expect in general an expansion of the form $F_{0}(\la) = F_{0}(\infty) + O(1/\sqrt{\la})$. The correction term $\la F'_{0}(\la)$ in \eqref{fundrelbis} is thus at most $O(1/\sqrt{\la})$ and, when $\la\rightarrow\infty$, we find the relation $\Sp^{*}=\frac{2}{N}\Sg^{*}$ used in Section \ref{Sec3}.

\noindent\ ii) We are going to focus on the specific example of the $\alp$-corrected Schwarzschild-$\AdS_{5}$ geometry. In this case, it turns out that the large $\la$ expansion is of the form
\be\label{Sexpla}\begin{split}
\Sg^{*}(\la) &= N^{2}\bigl(F_{0,0} + \la^{-3/2}F_{0,3/2} + O(\la^{-2})\bigr)\, ,\\\Sp^{*}(\la) &= N\bigl( f_{0} + \la^{-3/2}f_{3/2} + O(\la^{-2})\bigr)\, .
\end{split}\ee
Equation \eqref{frc2} then yields $f_{0} = 2 F_{0,0}$, which is the relation that we have already checked in Section \ref{Sec3}, together with the new constraint
\be\label{conseqf34} f_{3/2} = \frac{1}{2}  F_{0,3/2}\, .\ee
This is the relation that we are going to check below.

\noindent\ iii) At finite $\la$, the relation between $\Sp^{*}$ and $\Sg^{*}$ is no longer a simple proportionality, but \eqref{frc2} can always be integrated to find the on-shell supergravity action $\Sg^{*}(\la)$ from the on-shell brane action $\Sp^{*}(\la)$. To see this, one can, for example expand both sides of \eqref{frc2} at small $\la$ and check that the relation fixes the expansion to all orders.

\subsection{\label{UndSec} Schwarzschild-$\AdS_{5}$ to leading order}

We start by briefly reviewing the analysis at leading order. This analysis was already presented in \cite{FerRov} and also follows from the general discussion of Section 3. All we want here is to set-up the notations in a way convenient to the discussion of the $\alp$ corrections. In particular, we shall use a different  coordinate system than in \cite{FerRov}.

The leading order metric reads
\be\label{metAdSS}
	\d s_{0}^{2} = \frac{L^{2}}{u^{2}} \frac{\d u^{2}}{h(u,\alpha_{0})} + \frac{u^{2}}{a^{2}} \Big( h(u,\alpha_{0}) \d t^{2} + a^{2} \d \Omega_{3}^{2}\Big)\, ,
\ee
where the function $h(u,\alpha_{0})$ is given by
\be\label{defh}
	h(u,\alpha_{0}) = 1 + \frac{L^{2}}{u^{2}} \Big( 1+ \frac{(\alpha_{0}^{2}-1)L^{2}}{4 \alpha_{0}^{2} u^{2}} \Big)\, .
\ee
The parameters in the problem are $a$, the radius of the three-sphere $\text{S}^{3}$ on the boundary and the inverse temperature $\beta$.\footnote{Of course, $a$ is just a scale and we could set $a=1$, but we find it convenient to keep $a$ explicitly.} One should thus see $\alpha_{0}$, $0\leq\alpha_{0}\leq 1$, as being a function of these two parameters, such that
\be\label{tempAdSS}
	\beta = \pi a \sqrt{2 \alpha_{0} (1-\alpha_{0})}\, .
\ee
The range of the coordinate $u$ is $[\uh,+\infty[$, where the ``horizon'' is at  
\be\label{rhAdSS}
	\uh = \sqrt{\frac{1-\alpha_{0}}{2\alpha_{0}}}L\, .
\ee
Finally, let us note that the physically relevant root of the equation \eqref{tempAdSS} is given by
\be\label{azeroform} \alpha_{0} = \frac{1}{2}\biggl(1-\sqrt{1-\frac{2\beta^{2}}{\pi^{2}a^{2}}}\biggr)\ee
and corresponds to the large, stable Schwarzschild-AdS black hole.

Up to counter-terms, the free energy at leading order reads
\be\label{freeAdSS}
	F = \frac{N^{2}}{16 a} \frac{4\alpha_{0}-1}{\alpha_{0}^{2}}\, \cdotp
\ee
The D3 brane action for a world-volume of constant $u$ is
\be\label{probeAdSS}
	\Sp = \frac{2\pi^{2} \beta \tau_{3}u^{4}}{a}\Bigl(\sqrt{h(u,\alpha_{0})} -1+ \frac{\uh^{4}}{u^{4}} \Bigr) + s\, ,
\ee
where the constant $s$ is fixed as usual by the condition that asymptotically close to the boundary, $\Sp$ reduces to a counterterm, see Section \ref{sec3isop}. Using this prescription, we find that $s$ is given by
\be\label{vals}
	s =   \frac{\pi^{2}\beta \tau_{3}L^{4}}{a} \frac{4\alpha_{0}-1}{4\alpha_{0}^{4}}  + S_{\text{CT}} \, \cdotp 
\ee
Using \eqref{tau3AdS5} for the value of the tension $\tau_{3}$, the minimum of $\Sp$ is
\be\label{minproAdSS}
	\Sp^{*} = \frac{N \beta}{8 a} \frac{4 \alpha_{0}-1}{\alpha_{0}^{2}} + S_{\text{CT}}\, ,
\ee
consistently with \eqref{freeAdSS} and $\Sp^{*}=(2/N)\Sg^{*}$.


%
\subsection{On $\alp$-corrections}
\subsubsection{Relevant $\alp$-corrections to supergravity}\label{alpSec}

The type IIB supergravity action is corrected in string theory by higher derivative terms. The expansion parameter is $\alp/L^{2}\sim (\ls/L)^{2}$, where $L$ is the typical length scale of the background geometry. For us, $L$ is the $\AdS_{5}$ scale and the expansion parameter is simply
\be\label{lainfex65} \Bigl(\frac{\ls}{L}\Bigr)^{2} = \frac{2\pi}{\sqrt{\la}}\,\cdotp
\ee
These corrections have been extensively studied in the literature \cite{GZ,PZ,GW,FPSS,GKT}. At leading non-trivial order, it turns out that they are proportional $\ls^{6}$ and take the form
\be\label{gencorr}
	\Sg = \Sugra  -\frac{\zeta(3)\ls^{6}}{2^{10}\pi^{3} G_{10}} \int\! \d^{10}x  \sqrt{g}\,  e^{-3\phi/2} W\, .\ee
In this formula, $\phi$ is the dilaton fluctuation, normalized such that its kinetic term is
\be\label{scal}
	\frac{1}{16\pi G_{10}} \int \d^{10} x \sqrt{g}\ \frac{1}{2} g^{MN} \del_{M} \phi \del_{N}\phi\ee
and $g_{MN}$ is the Einstein frame metric, which is related to the string frame metric $G_{MN}$ by
\be\label{stringeinstein}
	G_{MN} = e^{\phi/2} g_{MN}\, .
\ee
Moreover, $W$ is a scalar, commonly called the ``$R^{4}$-term,'' constructed out of four powers of the Riemann curvature tensor. We are interested in cases where the Einstein frame metric is of the form
\be\label{genmet111} \d s^{2}_{10} = \d s^{2} + f \d \Omega^{2}_{5}\, ,\ee
where $\d s^{2}$ is a metric on a five-manifold $M$ and $f$ is some function on $M$. In this case, $W$ can be expressed as \cite{GKT,PT}
\be\label{WM}
	W = C^{c_{1}a_{1}a_{2}c_{2}}C_{d_{1}a_{1}a_{2}d_{2}}C_{c_{1}}^{\ b_{1}b_{2}d_{1}}C^{d_{2}}_{\ b_{1}b_{2}c_{2}} 
	+ \frac{1}{2} C^{c_{1}c_{2}a_{1}a_{2}}C_{d_{1}d_{2}a_{1}a_{2}}C_{c_{1}}^{\ b_{1}b_{2}d_{1}}C^{d_{2}}_{\ b_{1}b_{2}c_{2}}\, ,
\ee
where $1\leq a_{1},a_{2},b_{1}, b_{2},c_{1},c_{2},d_{1},d_{2} \leq 5$ and $C_{abcd}$ is the Weyl tensor for $\d s^{2}$. Additional corrections to the supergravity action, involving in particular the Ramond-Ramond forms, also exist, but do not affect our discussion (see \cite{GKT,MPS} and references therein).

\subsubsection{The $\alp$-corrected Schwarzschild-$\AdS_{5}$ geometry} \label{alpgen}

The new action \eqref{gencorr} yields the $\alp$-corrected equations of motion for the metric $g$ and the dilaton $\phi$. We are interested in the associated deformation of the Schwarzschild-$\AdS_{5}$ geometry \eqref{metAdSS}. This problem was first studied in \cite{Land}. In terms of the conveniently defined deformation parameter
\be\label{ve}
	\ve = \frac{\zeta(3)}{2^{6}\pi^{2}} \frac{\ls^{6}}{L^{6}} = \frac{\pi\zeta(3)}{8\la^{3/2}}\, \cvp
\ee
and the differential form
\be\label{omdefsol1}
\omega = \frac{1}{a} u^{3}\d u\wedge\omega_{\text{S}^{3}}\wedge\d t\, ,\ee
the solution reads, to leading non-trivial order in $\eta$,
\begin{align}\label{paramet}
	\d s^{2} &= \frac{L^{2}}{u^{2}} \frac{e^{A(u)} \d u^{2}}{h(u,\alpha)} + \frac{u^{2}}{a^{2}} \Big( h(u,\alpha) e^{B(u)} \d t^{2} + a^{2} e^{C(u)} \d \Omega_{3}^{2} \Big) + L^{2} e^{-3C(u)/5} \d \Omega_{5}^{2}
	\, , \\\label{F534}
	F_{5}  &=  4i \Big( e^{\frac{A(u)+B(u)}{2}+ 3C(u)} \omega  +i  L^{4}\omega_{\text{S}^{5} } \Big)\, .
\end{align}
The functions $A$ and $B$ are given in terms of $C$ by \cite{Land}
\begin{align}
\begin{aligned}\label{solLand}
A(u) &=C(u) + \frac{5 \ve L^{4}}{4\alpha^{2} h(u,\alpha) u^{4}} \biggl[ -\frac{(\alpha-15)(1+\alpha)^{3}}{(\alpha-1)^{2}}\\ &\hskip 4.3cm +
	 \frac{(\alpha^{2}-1)^{3}}{\alpha^{4}} \frac{L^{8}}{u^{8}} \Bigl( \frac{9}{2} + \frac{4L^{2}}{u^{2}} + \frac{57}{64} \frac{\alpha^{2}-1}{\alpha^{2}} \frac{L^{4}}{u^{4}}
	 \Bigr)
	\biggr] \, ,\end{aligned}\\
\label{solLand2}
\begin{aligned}
	B(u) &= C(u) + \frac{5 \ve L^{4}}{4 \alpha^{2} h(u,\alpha) u^{4}} \biggl[ \frac{(\alpha-15)(1+\alpha)^{3}}{(\alpha-1)^{2}}\\  &\hskip 4.3cm -\frac{(\alpha^{2}-1)^{3}}{\alpha^{4}} \frac{L^{8}}{u^{8}} \Bigl( \frac{3}{2} + \frac{L^{2}}{u^{2}} + \frac{9}{64} \frac{\alpha^{2}-1}{\alpha^{2}} \frac{L^{4}}{u^{4}}
	\Bigr)
\biggr]\, .\end{aligned}\end{align}
%
Let us note that the combination
\be\label{combABC} A(u)+B(u)-2C(u) = -\frac{15\eta}{4}\frac{(1-\alpha^{2})^{3}}{\alpha^{6}}\frac{L^{12}}{u^{12}}\ee
simplifies nicely. The function $C$ itself is not known explicitly, but it satisfies the second order differential equation
\be\label{gammafnt}
	\Bigl(  \frac{u^{5}}{L^{3}} h(u,\alpha) C'(u) \Bigr)' - \frac{32 u^{3}}{L^{3}} C(u) - \frac{225 \ve}{256} \frac{(1-\alpha^{2})^{4}}{\alpha^{8}} \frac{L^{13}}{u^{13}}  = 0 \, ,
\ee
where the prime denotes the derivative with respect to $u$. Imposing that the metric remains $\AdS_{5}$ at large $u$, we find from this equation that
\be\label{Casymp} C(u) \underset{r\rightarrow\infty}{\sim}\frac{c}{u^{8}}\ee
for some constant $c$. This is all what we shall need to know about $C$. An explicit expression for the dilaton $\phi(u)$ can also be found in \cite{Land}, but we won't use it.

As in the undeformed case reviewed in \ref{UndSec}, the parameters in the problem are the radius $a$ and the inverse temperature $\beta$. The generalization of \eqref{tempAdSS} reads
\be\label{TBHcorr}
	\beta = a\pi \sqrt{2 \alpha (1-\alpha)} \Big( 1- \frac{5(1+\alpha)^{3}(3-5\alpha)}{(1-\alpha)^{3}}\eta  \Big)\, ,\ee
which fixes $\alpha$ as a function of $a$ and $\beta$. Explicitly, the function $\alpha$ is given in terms of the function $\alpha_{0}$ defined in \eqref{azeroform} by
\be\label{alphaalpha0rel} \alpha = \alpha_{0}+\frac{5\beta^{2}}{\pi^{2}a^{2}}\frac{(1+\alpha_{0})^{3}(3-5\alpha_{0})}{(1-\alpha_{0})^{3}(1-2\alpha_{0})}\eta\, .\ee
Finally, the range of the coordinate $u$ is $[\uh,+\infty[$, with 
\be\label{horcorr}
	\uh = \sqrt{\frac{1-\alpha}{2\alpha}} L \, .
\ee
\subsubsection{Relevant $\alp$-corrections to the D3-brane action}\label{D3corrSec}

The leading-order D3-brane action is given by \eqref{probegenact} for $d=4$ and is proportional to $1/\alp^{2}$. The first $\alp$-corrections to this action have been studied in \cite{BBG}. The leading correction is $O(1)$ and the next-to-leading order is $O(\alp^{2})$,
\be\label{corrprobe}
	\delta\Sp = \delta \DBI + \delta \CS + O (\alp^{2})\, ,
\ee
where 
\begin{align}\label{corrDBI}
	\delta \DBI &= \frac{1}{16 \pi^{2}} \int\Re \big[  \log \big( \eta_{\text D}(\tau)\big) \tr \big(  \R \wedge \star \R -i \R \wedge \R  \big) \big]\, , \\
	\label{corrCS}
	\delta \CS &= -i \int \bigl( C_{0} + C_{2} \bigr) \wedge \Omega 
\end{align}
yield the deformations to the DBI and CS parts of the action respectively. We have denoted by $\eta_{\text D}$ the Dedekind function, by $\tau=(C_{0}+i e^{-\phi})/\la$ the axion-dilaton field and by $\R$ a two-form-valued matrix built using the pullback $\Rd$ of the Riemann tensor on the brane worldvolume as
\be\label{Rdetails}
	\R^{k}_{\ l} = \frac{1}{2} \Rd^{k}_{\ lij} \d x^{i} \wedge \d x^{j}\, .
\ee
The quantity $\Omega$ is a sum of differential forms, that can be computed from the Dirac roof genus. Since in the background we are studying, $C_{0}=C_{2}=0$, its detailed form will not be needed.

\subsection{The $\alp$-corrected on-shell actions}

We are now ready to evaluate both the on-shell gravitational action and the on-shell probe action, taking into account the $\alp$ corrections, in order to check our fundamental formula \eqref{fundrelbis}. The calculation for the on-shell gravitational action has already been done long ago in \cite{GKT,Land}, so the new part that we present is really the evaluation of the probe action. However, it is very interesting to explain both calculations in parallel. This will highlight some crucial differences in the way the $\alp$ corrections enter on both sides and underline the very non-trivial nature of the final match of the results, consistently with \eqref{fundrelbis}.

\subsubsection{The $\alp$-corrected on-shell gravitational action}

To evaluate $\Sg^{*}$, we plug the corrected geometry reviewed in \ref{alpgen} into \eqref{gencorr}. It is well-known \cite{GKT,PT} that, to leading non-trivial order in $\ve$, the supergravity action $\Sugra$ evaluated on this corrected geometry matches with the supergravity action evaluated on the undeformed geometry. The full $\alp$ corrections to $\Sg^{*}$ thus come from the evaluation of the $R^{4}$ term \eqref{WM}. Since this term is already $O(\ve)$, it is clear that we only need the undeformed geometry to make the calculation. The details of the corrected background presented in \ref{alpgen} thus turn out to be totally irrelevant here. Modulo the usual counterterms, this yields \cite{GKP,Land}
\be\label{freeAdSScorr}
	\Sg^{*} = \beta\frac{N^{2}}{16 a} \frac{4\alpha_{0}-1}{\alpha_{0}^{2}} \biggl[  1 + \frac{15 \ve (1+\alpha_{0})^{4}}{(1-4\alpha_{0})(1-\alpha_{0})^{2}} \biggr]+\CT\, ,
\ee
where the function $\alpha_{0}$ is defined by \eqref{azeroform}. In the notation of \eqref{Sexpla}, using \eqref{ve} and \eqref{tempAdSS}, we get
\be\label{Ffform} F_{0,0} =\frac{\beta}{16 a}\frac{4\alpha_{0}-1}{\alpha_{0}^{2}}  \, \cvp\quad F_{0,3/2} = -\frac{15\pi^{5}\zeta(3)}{32}\frac{a^{3}}{\beta^{3}}\bigl(1+\alpha_{0}\bigr)^{4}\, .\ee
\subsubsection{The $\alp$-corrected on-shell probe action}

We now compute the minimum of the corrected probe action, which includes the terms \eqref{corrprobe}, in the deformed geometry \eqref{paramet}, \eqref{F534}. 

Let us first consider the correction terms \eqref{corrDBI} and \eqref{corrCS}. Since $C_{0}=C_{2}=0$, \eqref{corrCS} clearly vanishes. As for \eqref{corrDBI}, at leading non-trivial order in $\eta$, it is a priori enough to evaluate it on the undeformed background at $r=\rh$. By direct calculation, it turns out that this term actually vanishes on this undeformed background for any fixed value of $r$. The conclusion is that the full $\alp$ corrections to $\Sp^{*}$ come from the evaluation of the undeformed probe action \eqref{probegenact} on the deformed background. 

Note that this is the exact opposite of what happened for the evaluation of the gravitational action $\Sg^{*}$. For $\Sg^{*}$, we needed the $\alp$ corrections to the action evaluated on the undeformed geometry and the deformed background played no r\^ole. Now, for $\Sp^{*}$, we see that the $\alp$ corrections to the action play no r\^ole, all the non-trivial contributions coming from the deformed geometry presented in \ref{alpgen}.

This being said, we can evaluate $\Sp^{*}$, following in details the prescription explained in Section \ref{sec3isop}.

\paragraph{Step 1:} The first part of the discussion precisely mimics what we have done in \ref{sec3isop}. Since, as we have just explained, the corrections $\delta S_{\text{DBI}}$ and $\delta S_{\text{SC}}$ to the probe action vanish in the present example, the probe action can be written as
\be\label{SpSt2} \Sp(\Sigma) = \tau_{3}\Bigl(A(\Sigma) +i\int_{M_{\Sigma}}\! F_{5}\Bigr) + s\, .\ee
This formula slightly generalizes \eqref{SpSt}, taking into account that \eqref{dCd} is no longer valid when the $\alpha'$ corrections are present. Moreover, it is not difficult to show that the minimum value of the probe action will be obtained for a shrunken brane, as in Section \ref{sec3isop}. This can be understood by using the spherical symmetry of the metric and by considering the potential felt by a brane at constant $u$, which is an infinitesimal perturbation of the potential felt by the brane in the undeformed case.\footnote{We have not tried to derive a generalization of the isoperimetric inequality \eqref{isopineq} which would apply in more general $\alpha'$-corrected background, but it seems clear that such generalizations should exist.} We thus obtain
\be\label{Sps34} \Sp^{*} = s\, .\ee

The constant $s$ is fixed as usual. We introduce the worldvolume $\Sigma_{\epsilon}$, given by the equation $r=\epsilon$, where $\epsilon$ is a regularizing parameter and $r$ the Fefferman-Graham coordinate. This coordinate is such that the bulk metric \eqref{paramet} takes the form \eqref{PEmet}, \eqref{FGexp} where, presently, the boundary metric is
\be\label{BMhere} \bar g = \d t^{2} + a^{2}\d\Omega_{3}^{2}\, .\ee
It is straightforward to express $r$ in terms of the coordinate $u$ in \eqref{paramet}. We only need the expansion near the boundary $r=0$ and this yields
\begin{multline}\label{Sepsu} \Sigma_{\epsilon}:\ u = u_{\epsilon}= \frac{a}{\epsilon}\biggl[ 1 -\frac{L^{2}\epsilon^{2}}{4a^{2}}\\
+\biggl(\frac{(1-\alpha_{0}^{2})L^{4}}{32a^{4}\alpha_{0}^{2}} + 
\frac{5\pi^{4}L^{4}}{8\beta^{4}}\frac{(1+\alpha_{0})^{3}(3-11\alpha_{0}+2\alpha_{0}^{2})}{1-2\alpha_{0}}\eta\biggr)\epsilon^{4}
+ O(\epsilon^{6}) \biggr]\, .
\end{multline}
We then require that $\Sp$ reduces to a counterterm action when evaluated on $\Sigma_{\epsilon}$. This yields
\be\label{scond34}\Sp^{*}=\lim_{\epsilon\rightarrow 0}\Bigl(-i\tau_{3}\int_{M_{\Sigma_{\epsilon}}}\! F_{5}+\CT(\Sigma_{\epsilon})\Bigr)\, .\ee

\paragraph{Step 2:} We evaluate the integral $\smash{-i\tau_{3} \int_{M_{\Sigma_{\epsilon}}}\! F_{5}}$, starting from \eqref{F534} and \eqref{omdefsol1}, to leading non-trivial order in $\eta$. The integrals over $t$ and $\text{S}^{3}$ are trivial. Using $\tau_{3}=\frac{N}{2\pi^{2}L^{4}}$, see \eqref{char00}, we then get
\be\label{int444} -i\tau_{3} \int_{M_{\Sigma_{\epsilon}}}\! F_{5} = \frac{4N\beta}{aL^{4}}\Bigl(I_{1}+I_{2}+I_{3}\Bigr)\, ,\ee
where the integrals over $u$ are conveniently written as
\begin{align}
\label{I1def} I_{1} &= \int_{\uh}^{u_{\epsilon}}\! u^{3}\d u\, ,\\
\label{I2def} I_{2} &= \frac{1}{2}\int_{\uh}^{u_{\epsilon}}\!\bigl(A(u)+B(u)-2C(u)\bigr)u^{3}\d u\, ,\\
\label{I3def}
I_{3} &= 4\int_{\uh}^{u_{\epsilon}}\! C(u)u^{3}\d u\, .
\end{align}
The integrals $I_{1}$ and $I_{2}$ can be immediately computed from the explicit expression \eqref{combABC}, recalling that $\alpha$ is given by \eqref{alphaalpha0rel} and that the bounds in the integrals are given by \eqref{horcorr} and \eqref{Sepsu}. In the small $\epsilon$ limit, we get
\begin{align}\label{I1val}
	 I_{1} &= \frac{a^{4}}{4 \epsilon^{4}} - \frac{a^{2}L^{2}}{4\epsilon^{2}} + \frac{L^{4}}{32 \alpha_{0}^{2}} \biggl[  4 \alpha_{0} -1 - \frac{15 \eta (1+\alpha_{0})^{3}(7\alpha_{0}-9)}{(1-\alpha_{0})^{2}} \biggr]+O(\epsilon)\, , \\
	\label{I23val}
	 I_{2} &= -\frac{15 L^{4}\eta}{4} \frac{(1+\alpha_{0})^{3}}{(1-\alpha_{0})\alpha_{0}^{2}}+O(\epsilon)\, .
\end{align}
In spite of the fact that the function $C(u)$ is not known explicitly, we can still compute the integral $I_{3}$ by integrating the differential equation \eqref{gammafnt} from $\uh$ to $u_{\epsilon}$ and by using the fall-off condition \eqref{Casymp}. This yields
\be\label{I4val}
	I_{3} = -\frac{75 L^{4} \eta}{128} \frac{(1+\alpha_{0})^{4}}{(\alpha_{0}-1)^{2}\alpha_{0}^{2}}+O(\epsilon)\, .\ee
Adding-up \eqref{I1val}, \eqref{I23val} and \eqref{I4val} we get, up to the divergent terms in $I_{1}$ which are counterterms,
\be\label{Spfinal}
	\Sp^{*} = \frac{N\beta}{8a} \biggl[  \frac{4\alpha_{0}-1}{\alpha_{0}^{2}}  - \frac{15 \pi^{4} a^{4}}{\beta^{4}} \bigl(1+\alpha_{0}\bigr)^{4}\eta \biggr]\, .
\ee
Using the relation \eqref{ve} between $\eta$ and $\lambda$ and comparing with the expansion of $\Sp^{*}$ in \eqref{Sexpla}, we therefore obtain
\be\label{finalterms}
	f_{0} =  \frac{\beta}{8 a} \frac{4 \alpha_{0}-1}{\alpha_{0}^{2}}  \, \cvp \quad f_{3/2} = - \frac{15 \pi^{5}\zeta(3)}{64} \frac{a^{3}}{\beta^{3}}  \bigl(1+\alpha_{0}\bigr)^{4}\, .
\ee
Comparing with \eqref{Ffform}, we see that the non-trivial prediction \eqref{conseqf34} of our fundamental relation \eqref{frc2} is beautifully satisfied.


%
\section{\label{CSec} Conclusion}

We have presented and studied in details a surprising relation \eqref{fundrel} between on-shell gravitational actions and on-shell probe actions, which was first proposed in \cite{fer1} and already explored in \cite{FerRov,McInnes}. We have emphasized that this relation is deeply rooted into the thermodynamical nature of gravity and as such should be extremely general. We have tested our ideas on several non-trivial examples, both in asymptotically flat and asymptotically AdS space-times.

The explicit tests we have provided, in particular in Section \ref{alphaSec}, are stringent. Let us briefly recapitulate the main ingredients that came into them.

Even to leading order in $\alpha'$, the match required several important consistency requirements, see \cite{FerRov} and Section \ref{Sec3}. Once the correct definition of the probe action was given, and the isoperimetric inequality established, it was still necessary to understand how a ``surface'' term like the probe action could match with a ``bulk'' term like the Einstein-Hilbert action. This was possible thanks to Stokes' theorem and the general relation \eqref{dCd} between the Ramond-Ramond field strength and the volume form of the bulk space-time. Finally, a precise numerical match between various combinations of string-theoretic quantities, Eq.\ \eqref{gammaform}, \eqref{D3prex}, \eqref{D1probe}, \eqref{D5probe}, \eqref{checkM2}, \eqref{checkM5}, had to be valid.

To the first non-trivial order in $\alpha'$, new highly non-trivial ingredients were needed. First, the fundamental formula no longer yields a simple proportionality between $\Sg^{*}$ and $\Sp^{*}$, but rather the differential relation \eqref{frc2}. From this, one gets specific relations between the coefficients in the large $\la$ expansions of $\Sg^{*}$ and $\Sp^{*}$, see Eq.\ \eqref{Sexpla} and \eqref{conseqf34}. Second, the match requires a very precise link between the integral of the $R^{4}$-terms \eqref{WM} over the undeformed space-time (which yields the $\alpha'$ corrections to the on-shell gravitational action) and the integral of $F_{5}$, Eq.\ \eqref{F534}, over the deformed space-time (which yields the $\alpha'$ corrections to the on-shell probe action). We were not able to find a simple direct proof from gravitational field equations of this fact, similar to the argument at leading order based on \eqref{dCd}. We just checked that it works from direct evaluation of the integrals. 

We believe that all this constitutes a very convincing check of the consistency of the general arguments that underly our fundamental relation \eqref{fundrel} and thus, in particular, of the thermodynamic nature of gravity.

One can imagine many lines of future developments. Let us briefly mention three of them.

In the strict thermodynamic limit where a purely classical description of gravity is valid, a fundamental consistency requirement is the existence of the isoperimetric inequality \eqref{isopineq}. Our work suggests that a whole new class of isoperimetric inequalities should exist for more general backgrounds, in which the notions of area and volume are replaced by the DBI and Chern-Simons terms in the brane probe action. Even more generally, these inequalities should generalize to higher-derivative gravitational theories. In view of the importance of isoperimetric inequalities in geometry, which is due, in particular, to their deep link with spectral theory, a research along these lines could be very fruitful.

In view of the extreme generality of our arguments, it should be possible to derive and check versions of our fundamental relation \eqref{fundrel} in many different contexts. A particularly interesting framework is provided by the ``bubbling geometries'' obtained in \cite{bubbling1,bubbling2,bubbling3}.\footnote{We would like to thank the referee for pointing out some of these references to us.} For instance, one can consider geometries dual to half-BPS Wilson loops of the $\nn=4$ super Yang-Mills theory in large representations of the gauge group $\text{SU}(N)$, with Young tableaux containing of the order of $N^{2}$ boxes. The full back-reaction on the $\AdS_{5}\times\text{S}^{5}$ geometry must then be taken into account. The corresponding $\text{AdS}_{2}\times\text{S}^{2}\times\text{S}^{4}\times\Sigma$ solutions were constructed in \cite{bubbling3}. These solutions are parameterized by harmonic functions which themselves encode the large Young tableau of the Wilson loop representation \cite{Tranca}. We can then apply our fundamental relation \eqref{fundrel}, with $Q$ identified with the total fundamental string charge of the solutions, which equals the total number of boxes in the Young tableau. From the known relation between D3 and D5 branes attached to the Wilson loop contour and Wilson loop representations \cite{WilsonHolIrrep}, \eqref{fundrel} predicts the non-trivial equality between:\\
--- on the one hand, the on-shell probe actions of a D3 brane or a D5 brane wrapping $\text{AdS}_{2}\times\text{S}^{2}$ or $\text{AdS}_{2}\times\text{S}^{4}$ respectively, carrying $q$ units of fundamental string charge on $\text{AdS}_{2}$ and attached to the circular Wilson loop contour on the $\text{AdS}_{5}$ boundary;\\ --- and on the other hand, the variation of the on-shell supergravity action for the $\text{AdS}_{2}\times\text{S}^{2}\times\text{S}^{4}\times\Sigma$ solution, when the fundamental string charge is varied by $q$ units, which corresponds to adding a row (for D3 branes) or a column (for D5 branes) with $q$ boxes to the Young tableau.\\
We believe that several other non-trivial illustrations of \eqref{fundrel} could be found along similar lines.

Finally, an ambitious but very interesting question is to go beyond the strict thermodynamic limit of gravity. This is required, for example, to study the higher-derivative corrections of the M-brane backgrounds discussed in Section \ref{sec3Mbranes}, which are related to $1/N$ corrections. To work at finite $N$, one clearly needs an understanding of the ``microscopic'' definitions of the probe branes actions, along the lines of \cite{fer1} (see Appendix B). We believe that deep and unexpected consistency requirements in quantum gravity may be uncovered in this way.

\subsection*{Acknowledgments}

The work of F.F.\ is supported in part by the Belgian Fonds National de la Recherche Scientifique FNRS (convention IISN 4.4503.15, CDR grant J.0088.15 and MS grant) and the Advanced ARC project ``Holography, Gauge Theories and Quantum Gravity.''

The work of A.R.\ is supported in part by the DFG Transregional Collaborative Research Centre TRR 33 and the DFG cluster of excellence ``Origin and Structure of the Universe,'' the Belgian American Educational Foundation as well as the Fonds National Suisse, subsidies 200021-162796, and by the NCCR 51NF40-141869 ``The Mathematics of Physics'' (SwissMAP).

\appendix
\numberwithin{equation}{section}
\part*{Appendices}
\label{appendices}
\section{Conventions}\label{convApp}

In this Appendix we summarize all our conventions and give some useful formulas that are used in the body of the paper, paying particular attention to all  signs and factors of 2 and $\pi$. We work exclusively in the Euclidean. Otherwise, we use the conventions of Polchinski's standard string theory textbook \cite{Popol}.

The string length $\ls$ is related to $\alp$ by
\be\label{deflsapp}
	\ls^{2}= 2\pi\alp\, .
\ee
For a $(d+1)$-dimensional spacetime $\M$ with boundary $\del \M$, the Einstein-Hilbert action $\EH$ reads
\be
\label{defEH}
	\EH = - \frac{1}{16 \pi G_{d+1}} \int_{\M} \d^{d+1}x \sqrt{G} \, \big( R-2\La\big)\, .\ee
The Riemann curvature tensor is defined with the following sign convention:
\be\label{defRiemann}
	R^{\mu}_{\ \nu\rho\lambda} = \del_{\rho} \Gamma^{\mu}_{\lambda \nu} -  \del_{\lambda} \Gamma^{\mu}_{\rho \nu} + \Gamma^{\mu}_{\rho \kappa}\Gamma^{\kappa}_{\lambda\nu} - \Gamma^{\mu}_{\lambda \kappa}\Gamma^{\kappa}_{\rho\nu}\, ,
\ee
and the Ricci tensor is $R_{\mu\nu} = R^{\lambda}_{\ \mu\lambda\nu}$. For asympotically AdS spacetime $\M$, the cosmological constant $\La$ is related to the AdS ``radius'' $L$ by
\be\label{defL}
	L^{2} = -\frac{(D-1)(D-2)}{2 \La}\, \cdot
\ee
%

The Dirac-Born-Infled action $\DBI$ for a probe $p$-brane moving in $\M$ reads
\be\label{defDBI2}
	\DBI = \tau_{p} \int_{\Si} \d^{p+1}x\ e^{-\phi} \sqrt{\det\big[\P (G+B)+\ls^{2}F \big]}\, ,
\ee
where $\Si$ is the $(p+1)$-dimensional worldvolume of the brane and $\P$ is the pullback on $\Si$. The field $B$ is the Kalb-Ramond field, if any. $F$ is the field-strength associated to the worldvolume U(1) gauge potential $A$. $\phi$ is the dilaton fluctuation, and therefore $\tau_{p}$ is proportional to $\gs^{-1}$ (see below for explicit formulas in superstring theory and in M-theory).

The Chern-Simons action for the $p$-brane (also called the Wess-Zumino term) reads
\be\label{defCS2}
	\CS =- i \tau_{p} \int_{\Si}\sum_{k}\P(C_{k}) \wedge e^{\P(B)+\ls^{2}F}\, .
\ee
The factor of $i$ comes from the Euclidean signature. The sum over $k$ runs over all allowed value in the given supergravity theory.

The ten-dimensional Newton constant $G_{10}$ in superstring theory is
\be\label{G10app}
	G_{10} = \frac{\pi^{2}}{2}\ls^{8}\gs^{2}\, .
\ee
The tension $\tau_{p}$ for a D$p$-brane is given by
\be\label{tensionDpapp}
	\tau_{p} = \frac{1}{(2\pi)^{\frac{p-1}{2}}\ls^{p+1}\gs}\, \cdot
\ee
The eleven-dimensional Newton constant $G_{11}$ and the eleven-dimensional Planck length $\LPM$ in M-theory are given by
\be\label{G11app}
	G_{11} =\LPM^{9} = \sqrt{\frac{\pi^{5}}{2}} \gs^{3} \ls^{9}\, , \quad\quad \LPM = \frac{\pi^{5/18}}{2^{1/18}} \gs^{1/3}\ls\, .
\ee
The M2- and M5-brane tensions are
\be\label{M2M5tensionsapp}
	\Mtwo =  \frac{1}{\sqrt{2\pi}\, \gs \ls^{3}} \, \cvp \quad \quad \Mfive = \frac{1}{(2\pi)^{2}\gs^{2}\ls^{6}} \, \cdot
\ee
\section{\label{GeneralitiesSec} On D-brane probes in gauge theory}

For completeness, we very briefly review the precise gauge theory framework, developed in \cite{fer1}, which provides a solid conceptual background for the relation \eqref{hol2} in the context of large $N$ gauge theories. In particular, a gauge-theoretic proof of \eqref{fundrel} is given and its generalization to all orders in the $1/N$ expansion or even at finite $N$ is discussed.

\subsection{Generalities}

Let us consider a $\uN$, or $\suN$, gauge theory in $p+1$ space-time dimensions, for example the four dimensional $\nn=4$ super Yang-Mills theory or the pure Yang-Mills theory. It was shown in \cite{fer1,fer1bis} that it is possible to define, purely in gauge-theoretic terms, what is meant by the ``microscopic'' non-Abelian D-brane action $\mathscr A_{N,K}$ for $K$ D$p$-branes in the presence of $N$ other D$p$-branes. The fundamental property of the action $\mathscr A_{N,K}$, which has a $\uK$ gauge symmetry, is to compute the ratio $Z_{N+K}/Z_{N}$ of partition functions (or generating functionals) of the original $\text{U}(N+K)$ and $\text{U}(N)$ gauge theories. Precisely, if we denote collectively by $\Phi$ the field variables that enter in $\mathscr A_{N,K}$, and if we work in the Euclidean, we have the path integral formula
\be\label{fundrelmic} \frac{Z_{N+K}(\la)}{Z_{N}(\la)} = \int\!\mathcal D\Phi\mathcal D[\text{ghosts}]_{\uK}\, e^{-\mathscr A_{N,K}(\Phi) + s\psi}\, ,\ee
where ghosts and a gauge-fixing term $s\psi$ have been introduced to take care of the $\uK$ gauge symmetry. The action $\mathscr A_{N,K}$ has several  interesting properties, discussed in details in \cite{fer1}. Most notably:

\noindent (i) The set of fields $\Phi$ include scalar fields. These fields describe the motion of the D$p$-branes in an emergent space, transverse to the original $p+1$ dimensional space-time, which is identified with the holographic geometry dual to the gauge theory under consideration. This is true even when the gauge theory does not have any elementary scalar field in the Lagrangian and provides a very effective approach to derive the holographic description of gauge theories. For example, a holographic fifth dimension is explicitly seen to emerge in this way in the pure Yang-Mills theory in four dimensions \cite{fer1}.

\noindent (ii) The action $\mathscr A_{N,K}$ provides a precise tool to probe the holographic bulk dual \emph{locally}. The difficulty in defining local observables in the bulk is rigorously addressed in the construction of the action $\mathscr A_{N,K}$, by mapping the non--gauge invariance of the local coordinates to the non-standard equivariant gauge-fixing procedure which is crucially needed to define $\mathscr A_{N,K}$ \cite{fer1,ferequiv}. The full details of the bulk geometry, like the metric or form-fields, can be read off from $\mathscr A_{N,K}$ \cite{ferrelated}.

\noindent (iii) At large $N$ and fixed $K$, which is the so-called probe limit, the action $\mathscr A_{N,K}$ has an expansion of the form
\be\label{ANKexp} \mathscr A_{N,K} = \sum_{k\geq 0}N^{1-k}A_{K}^{(k)}\, ,\ee
for actions $A_{K}^{(k)}$ that are independent of $N$. In particular, the probe brane action is proportional to $N$ at large $N$, $\mathscr A_{N,K}\simeq N \smash{A_{K}^{(0)}}$.

\subsection{The leading large $N$ limit}\label{largeNSec}

The path integral formula \eqref{fundrelmic} greatly simplifies in the large $N$ limit \cite{fer1}. To leading order, its right-hand side can be straightforwardly evaluated via the saddle point approximation, since the action $\mathcal A_{N,K}$ is proportional to $N$ and the number of fields in the set $\Phi$ is $N$-independent. If we denote by $A_{K}^{*}$ the on-shell value of the leading term $\smash{A_{K}^{(0)}}$ in the expansion \eqref{ANKexp}, we get
\be\label{saddlefund} \int\!\mathcal D\Phi\mathcal D[\text{ghosts}]_{\uK}\, e^{-\mathscr A_{N,K}(\Phi) + s\psi} = e^{-N A_{K}^{*} + O(1)}\, .\ee
On the other hand, the large $N$, fixed $K$ limit of the ratio of partition functions can be obtained from the usual large $N$ expansion \eqref{Zexpgauge}. This yields
\be\label{ratiolargeN}\frac{Z_{N+K}}{Z_{N}} = e^{-2NK F_{0}+O(1)}\, .\ee
Comparing \eqref{saddlefund} and \eqref{ratiolargeN}, we get
\be\label{fundEqual} A_{K}^{*} = 2K F_{0}\, .\ee
Since the planar free energy obviously does not depend on $K$, \eqref{fundEqual} implies a trivial relation between the on-shell Abelian and non-Abelian D-brane actions,
\be\label{AKA1rel}A_{K}^{*}=KA_{1}^{*}\, .\ee
For this reason, it is enough to concentrate on the Abelian case $K=1$. Noting $A_{1}=A$, \eqref{fundEqual} takes the form
\be\label{fundEqual2} A^{*} = 2 F_{0}\, .\ee
This relation, derived for any value of the 't~Hooft's coupling $\la$, is equivalent to \eqref{hol2} or \eqref{fundrelbis}. The apparent discrepancy comes from the fact that the definition of the probe action given in \cite{fer1} is designed in such a way that the formula \eqref{fundrelmic} is valid at a fixed value of the 't~Hooft's coupling, whereas the natural string theory definition amounts to working at fixed string or gauge coupling. This subtlety is fully clarified in \ref{twodefSec} below.

\subsection{On $1/N$ corrections}\label{oneoverN}

The relation \eqref{fundrelmic} is valid for any finite $N$ and $K$ and thus can be used beyond the leading large $N$ approximation.

Let us expand both sides of Eq.\ \eqref{fundrelmic} in powers of $1/N$. Using \eqref{Zexpgauge}, the left hand side yields
\be\label{beyondlN} -\ln\frac{Z_{N+K}}{Z_{N}} = 2NKF_{0} + K^{2}F_{0} - \frac{2K}{N^{3}}F_{2} + O\bigl(N^{-4}\bigr)\, ,\ee
whereas the right-hand side can be written
\be\label{rhssub} -\ln\int\!\mathcal D\Phi\mathcal D[\text{ghosts}]_{\uK}\, e^{-N A_{K}^{(0)} - A_{K}^{(1)} - N^{-1}A_{K}^{(2)} - N^{-2}A_{K}^{(3)} - N^{-3}A_{K}^{(4)} + O(N^{-4}) + s\psi}\, .\ee
Equating \eqref{beyondlN} and \eqref{rhssub} yields some rather non-trivial constraints on the probe brane path integral. The one-loop contribution computed with the action $\smash{A_{K}^{(0)}}$, supplemented with the on-shell value of $\smash{A_{K}^{(1)}}$, must reproduce the genus zero contribution $F_{0}$; contributions at order $N^{-1}$ and $N^{-2}$, which involve up to three loops, must both cancel; the contribution at order $N^{-3}$ must reproduce the genus one term $F_{2}$; etc. These properties are direct consequences of \eqref{fundrelmic} but constitute highly non-trivial predictions from the dual gravitational perspective. 
\subsection{Two natural definitions of the probe action}\label{twodefSec}

The D-brane action $\mathscr A_{N,K}$ is defined so that Eq.\ \eqref{fundrelmic} is satisfied \cite{fer1}. On the left-hand side of this equation, both $Z_{N+K}$ and $Z_{N}$ are evaluated at the same 't~Hooft's coupling $\la$. This implies that the gauge coupling constants $g^{2}$ are not the same in the $\text{U}(N+K)$ and $\text{U}(N)$ theories. 

It is equally natural to work at fixed gauge coupling and define a different D-brane action $\tilde{\mathscr A}_{N,K}$ such that 
\be\label{fundrelmic2} \frac{Z_{N+K}(g^{2})}{Z_{N}(g^{2})} = \int\!\mathcal D\Phi\mathcal D[\text{ghosts}]_{\uK}\, e^{-\tilde{\mathscr A}_{N,K}(\Phi) + s\psi}\, ,\ee
where, now, the partition functions on the left-hand side are evaluated for the same gauge coupling $g^{2}$. The construction of \cite{fer1} can be trivially adapted to this case. At leading order in $N$, the two brane actions $\mathscr A_{N,K}(\Phi)$ and $\tilde{\mathscr A}_{N,K}(\Phi)$ simply differ by a $\Phi$-independent, but coupling-dependent, constant. If the action of the gauge theory is written in the usual single-trace form
\be\label{Lag} S = \frac{1}{g^{2}}\int\!\d x\, \tr L\, ,\ee
the precise relation reads
\be\label{AArel} \mathscr A_{N,K}(\Phi) = \tilde{\mathscr A}_{N,K}(\Phi) + \frac{K}{\la}\int\!\d x\,\bigl\langle\tr L\bigr\rangle_{\text{pl.}} + O\bigl(N^{0}\bigr)\, ,\ee
where $\langle\tr L\rangle_{\text{pl.}}$ is the planar expectation value of the Lagrangian. This expectation value was always explicitly included in \cite{fer1} and \cite{fer1bis} (for example, it corresponds to the term in $\langle S_{N}(V)\rangle$ in Eq.\ (3.13) of \cite{fer1}, and similar terms in other equations). 

It is easy to check that the relation \eqref{AArel} ensures the consistency between \eqref{fundrelmic} and \eqref{fundrelmic2}. Indeed, noting that a given gauge coupling $g^{2}$ corresponds to the 't~Hooft's coupling $\la$ and $\frac{N}{N+K}\la$ in the $\text{U}(N+K)$ and $\text{U}(N)$ gauge theories respectively, and using the fact that
\be\label{Zder} \frac{\partial \ln Z_{N}}{\partial \la} = 
\frac{N}{\la^{2}}\int\!\d x\,\bigl\langle\tr L\bigr\rangle\, ,\ee
we get, to leading order at large $N$,
\begin{align}
\ln\frac{Z_{N+K}(g^{2})}{Z_{N}(g^{2})}& = \ln\frac{Z_{N+K}\bigl(\la\bigr)}{Z_{N}\bigl(\frac{N}{N+K}\la\bigr)} = \ln\frac{Z_{N+K}(\la)}{Z_{N}(\la)}
+\frac{K}{N}\la\frac{\partial\ln Z_{N}}{\partial\la} + O\bigl(N^{0}\bigr)\\
\label{ideforrel}
&=\ln\frac{Z_{N+K}(\la)}{Z_{N}(\la)} + 
\frac{K}{\la}\int\!\d x\,\bigl\langle\tr L\bigr\rangle_{\text{pl.}}
+ O\bigl(N^{0}\bigr)\, .
\end{align}
Since, at large $N$, the path integral representations \eqref{fundrelmic2} and \eqref{fundrelmic} show that the ratios of partition functions can be evaluated in terms of the on-shell brane actions, \eqref{ideforrel} is equivalent to
\be\label{ideforrel2} -\tilde{\mathscr A}_{N,K}^{*} = -\mathscr A_{N,K}^{*} 
+ 
\frac{K}{\la}\int\!\d x\,\bigl\langle\tr L\bigr\rangle_{\text{pl.}}
+ O\bigl(N^{0}\bigr)\, ,\ee
which follows from \eqref{AArel} by going on-shell.

In string theory, the gauge coupling constant is related to the string coupling constant, and the most natural choice is to keep the string coupling fixed when branes are added. \emph{The usual probe brane action in string theory is thus identified with the gauge theory action $\tilde{\mathscr A}_{N,K}$.} To leading order at large $N$, and in the Abelian case $K=1$, we have noted this action $\Sp$ in the main text,
\be\label{Sbdef} \tilde{\mathscr A}_{N,1} = \Sp + O\bigl(N^{0}\bigr)\, .\ee

We can now easily relate the on-shell value of the probe brane action to the planar free energy $F_{0}$. Combining \eqref{fundEqual2} with \eqref{ideforrel2} indeed yields
\be\label{derf3} \Sp^{*} = 2NF_{0} + 
\frac{1}{\la}\int\!\d x\,\bigl\langle\tr L\bigr\rangle_{\text{pl.}}\, .\ee
Since, in the planar limit, \eqref{Zder} is equivalent to 
\be\label{Zderpl} N\frac{\partial F_{0}}{\partial\la} = \frac{1}{\la^{2}}\int\!\d x\,\bigl\langle\tr L\bigr\rangle_{\text{pl.}}\, ,\ee
we see that \eqref{derf3} is equivalent to the fundamental relation \eqref{fundrelbis}.

%
%

%

%
\end{document}